\documentclass[useAMS,usenatbib]{mn2e}
\usepackage{times}
\usepackage{psfig}

\newcommand{\eref}[1]{(\ref{#1})}
\renewcommand{\d}{{\rm d}}
\newcommand{\ion}[2]{#1$\,${\sc #2}}
\newcommand{\ol}{\Omega_\Lambda}
\newcommand{\om}{\Omega_{\rm M}}
\newcommand{\ok}{\Omega_k}
\newcommand{\lya}{Ly$\alpha$}
\newcommand{\lyb}{Ly$\beta$}
\newcommand{\lyg}{Ly$\gamma$}
\newcommand{\kms}{~km~s$^{-1}$}
\newcommand{\cms}{~cm~s$^{-1}$}
\newcommand{\tem}{t_{\rm em}}
\newcommand{\tobs}{t_{\rm obs}}
\newcommand{\zdot}{$\dot z$}
\newcommand{\vdot}{\dot v}
\newcommand{\zd}{z_{\rm D}}
\newcommand{\zobs}{z_{\rm obs}}
\newcommand{\zqso}{z_{\rm QSO}}
\newcommand{\ti}{t_{\rm int}}
\newcommand{\vc}{\frac{v}{c}}
\newcommand{\ac}{\frac{a}{c}}
\newcommand{\p}{^\prime}
\newcommand{\dpa}{d_{\rm px}}
\newcommand{\vt}{v_{\rm t}}
\newcommand{\ar}{a_{\rm r}}
\newcommand{\vp}{v_{\rm p}}
\newcommand{\ap}{a_{\rm p}}
\newcommand{\tp}{\theta_{\rm p}}
\newcommand{\sv}{\sigma_v}
\newcommand{\nqso}{N_{\rm QSO}}
\newcommand{\np}{N_{\rm phot}}
\newcommand{\lcdm}{$\Lambda$CDM}

\title[Cosmic dynamics]{Cosmic dynamics in the era of
  Extremely Large Telescopes}
\author[J.~Liske et~al.]{J.~Liske,$^1$\thanks{E-mail: jliske@eso.org}
A.~Grazian,$^2$ E.~Vanzella,$^3$ M.~Dessauges,$^4$ M.~Viel,$^{3,5}$ 
L.~Pasquini,$^1$\newauthor
M.~Haehnelt,$^5$ S.~Cristiani,$^3$ F.~Pepe,$^4$ G.~Avila,$^1$ 
P.~Bonifacio,$^{6,3}$ F.~Bouchy,$^{7,8}$\newauthor
H.~Dekker,$^1$ B.~Delabre,$^1$ S.~D'Odorico,$^1$ V.~D'Odorico,$^3$
S.~Levshakov,$^9$ C.~Lovis,$^4$\newauthor
M.~Mayor,$^4$ P.~Molaro,$^3$ L.~Moscardini,$^{10,11}$ M.T.~Murphy,$^{5,12}$
D.~Queloz,$^4$ P.~Shaver,$^1$\newauthor
S.~Udry,$^4$ T.~Wiklind$^{13,14}$ and S.~Zucker$^{15}$\\
$^1$European Southern Observatory, Karl-Schwarzschild-Str.~2, 85748
Garching, Germany\\
$^2$INAF -- Osservatorio Astronomico di Roma, via di Frascati 33, 00040
Monteporzio Catone (Roma), Italy\\
$^3$INAF -- Osservatorio Astronomico di Trieste, Via Tiepolo 11, 34143
Trieste, Italy\\
$^4$Observatoire de Gen{\`e}ve, 51 Ch.~des Maillettes, 1290 Sauverny,
Switzerland\\
$^5$Institute of Astronomy, University of Cambridge, Madingley Road,
Cambridge CB3 0HA\\
$^6$CIFIST Marie Curie Excellence Team, GEPI, Observatoire de Paris,
CNRS, Universit\'e Paris Diderot, Place Jules Janssen 92190 Meudon,
France\\
$^7$Laboratoire d'Astrophysique de Marseille, Traverse du Siphon,
13013 Marseille, France\\
$^8$Observatoire de Haute-Provence, 04870 St Michel l'Observatoire, France\\
$^9$Department of Theoretical Astrophysics, Ioffe Physico-Technical
Institute, 194021 St.~Petersburg, Russia\\
$^{10}$Dipartimento di Astronomia, Universit{\`a} di Bologna, via Ranzani 1,
40127 Bologna, Italy\\
$^{11}$INFN -- National Institute for Nuclear Physics, Sezione di
Bologna, viale Berti Pichat 6/2, 40127 Bologna, Italy\\
$^{12}$Centre for Astrophysics \& Supercomputing, Swinburne
University of Technology, Hawthorn, VIC 3122, Australia\\
$^{13}$Space Telescope Science Institute, 3700 San Martin Drive,
Baltimore MD 21218, USA\\
$^{14}$Affiliated with the Space Sciences Department of the European
Space Agency\\
$^{15}$Department of Geophysics and Planetary Sciences,
Raymond and Beverly Sackler Faculty of Exact Sciences,
Tel Aviv University, Tel Aviv 69978, Israel}

\begin{document}

\date{Accepted
...... Received .....}

\pagerange{\pageref{firstpage}--\pageref{lastpage}} \pubyear{2007}

\maketitle

\label{firstpage}

\begin{abstract}
The redshifts of all cosmologically distant sources are expected to
experience a small, systematic drift as a function of time due to the
evolution of the Universe's expansion rate. A measurement of this
effect would represent a direct and entirely model-independent
determination of the expansion history of the Universe over a redshift
range that is inaccessible to other methods. Here we investigate the
impact of the next generation of Extremely Large Telescopes on the
feasibility of detecting and characterising the cosmological redshift
drift. We consider the Lyman $\alpha$ forest in the redshift range $2
< z < 5$ and other absorption lines in the spectra of high redshift
QSOs as the most suitable targets for a redshift drift experiment.
Assuming photon-noise limited observations and using extensive Monte
Carlo simulations we determine the accuracy to which the redshift
drift can be measured from the \lya\ forest as a function of
signal-to-noise and redshift. Based on this relation and using the
brightness and redshift distributions of known QSOs we find that a
$42$-m telescope is capable of unambiguously detecting the redshift
drift over a period of $\sim$$20$~yr using $4000$~h of observing
time. Such an experiment would provide independent evidence for the
existence of dark energy without assuming spatial flatness, using any
other cosmological constraints or making any other astrophysical
assumption.
\end{abstract}


\section{Introduction}
\label{intro}
The universal expansion was the first observational evidence that
general relativity might be applicable to the Universe as a
whole. Since \citeauthor{Hubble29}'s (\citeyear{Hubble29}) discovery
much effort has been invested into completing the basic picture of
relativistic cosmology. The central question is: what is the
stress-energy tensor of the Universe? Assuming homogeneity and
isotropy reduces this question to: what is the mean density and
equation of state of each mass-energy component of the Universe? Since
these parameters determine both the evolution with time and the
geometry of the metric that solves the Einstein equation, one can use
a measurement of either to infer their values. Over the past decade
the successes on this front have reached their (temporary)
culmination: observations of the Cosmic Microwave Background
\citep[CMB;][]{Spergel03,Spergel07}, type Ia supernovae
\citep[SNIa;][]{Riess04,Astier06}, the large-scale galaxy distribution
\citep{Peacock01,Cole05,Eisenstein05} and others now provide answers
of such convincing consistency and accuracy that the term `precision
cosmology' is now commonplace \citep[e.g.][]{Primack05}.

By far the most unexpected result of this campaign was the discovery
that the expansion of the Universe has recently begun accelerating
\citep{Riess98,Perlmutter99}. The physical reason for this
acceleration is entirely unclear at present. Within relativistic
cosmology it can be accommodated by modifying the stress-energy tensor
to include a new component with negative pressure. In its simplest
incarnation this so-called dark energy is the cosmological constant
$\Lambda$ \citep*[e.g.][]{Carroll92}, i.e.\ a smooth, non-varying
component with equation of state parameter $w = -1$. It could also be
time variable \citep[e.g.][]{Overduin98}, either because of a specific
equation of state (xCDM, e.g.\ \citealp{Turner97}; phantom energy,
\citealp{Caldwell02}) or because the component consists of a dynamical
scalar field evolving in a potential \citep[e.g.][]{Ratra88}. More
generally it may be an inhomogeneous, time varying component with $-1
\la w \la 0$ (where $w$ may also vary with time), sometimes called
quintessence \citep*[e.g.][]{Caldwell98}, or even a component with an
exotic equation of state (e.g.\ Chaplygin gas,
\citealp*{Kamenshchik01}). Alternatively, instead of modifying the
stress-energy tensor one can also modify gravity itself to explain the
acceleration (\citealp*[e.g.][]{Deffayet02};
\citealp{Freese02,Carroll05}). Although all current observations are
consistent with a cosmological constant \citep*[e.g.][]{Seljak06} many
more unfamiliar models are not ruled out
\citep[e.g.][]{Capozziello05}.

Probably the best way to probe the nature of the acceleration is to
determine the expansion history of the Universe
\citep{Linder03b,Seo03}. Observables that depend on the expansion
history include distances and the linear growth of density
perturbations \citep{Linder03a,Linder05}, and so SNIa surveys, weak
lensing \citep{Heavens03,Jain03} and baryon acoustic oscillations in
the galaxy power spectrum \citep[BAO;][]{Seo03,Wang06} are generally
considered to be excellent probes of the acceleration.

In practice, however, extracting information on the expansion history
from weak lensing and BAO requires a prior on the spatial curvature, a
detailed understanding of the linear growth of density perturbations
and hence a specific cosmological model. Given the uncertain state of
affairs regarding the source of the acceleration, and given that even
simple parameterisations of dark energy properties can result in
misleading conclusions \citep{Maor01,Bassett04}, these are
conceptually undesirable features and several authors have pointed out
the importance of taking a cosmographic, model-independent approach to
determining the expansion history
\citep[e.g.][]{Wang05,John05,Shapiro06}. Using SNIa to measure
luminosity distances as a function of redshift is conceptually the
simplest experiment and hence appears to be the most useful in this
respect. The caveats are that distance is `only' related to the
expansion history through an integral over redshift and that one still
requires a prior on spatial curvature \citep{Caldwell04}.

Here we will revisit a method to directly measure the expansion
history that was first explored by \citet[][with an appendix by
McVittie]{Sandage62}. He showed that the evolution of the Hubble
expansion causes the redshifts of distant objects partaking in the
Hubble flow to change slowly with time. Just as the redshift, $z$, is
in itself evidence of the expansion, so is the change in redshift,
\zdot, evidence of its de- or acceleration between the epoch $z$ and
today. This implies that the expansion history can be determined, at
least in principle, by means of a straightforward spectroscopic
monitoring campaign.

The redshift drift is a direct, entirely model-independent measurement
of the expansion history of the Universe which does not require any
cosmological assumptions or priors whatsoever. However, the most
unique feature of this experiment is that it directly probes the
global {\em dynamics} of the metric. All other existing cosmological
observations, including those of the CMB, SNIa, weak lensing and BAO,
are essentially {\em geometric} in nature in the sense that they map
out space, its curvature and its evolution. Many of these experiments
also probe the dynamics of localised density perturbations but none
actually measure the {\em global} dynamics. In this sense the redshift
drift method is qualitatively different from all other cosmological
observations, offering a truly independent and unique approach to the
exploration of the expansion history of the Universe.

Following the original study by \citet{Sandage62}, the redshift drift
and its relevance to observational cosmology were also discussed by
\citet{McVittie65}, \citet{Weinberg72}, \citet{Ebert75},
\citet{Rudiger80}, \citet{Peacock99}, \citet{Nakamura99},
\citet{Zhu04}, \citet*{Corasaniti07} and \citet{Lake07}.
\citet{Lake81} gave equations expressing the deceleration and matter
density parameters, $q_0$ and $\om$, in terms of $z$, \zdot\ and
$\ddot z$. An excellent expos{\'e} of the equations relevant to
redshift evolution was also presented by \citet{Gudmundsson02}. In
addition, these authors investigated the redshift drift in the
presence of quintessence, while other non-standard dark energy models
were considered by \citet{Balbi07} and \citet{Zhang07}. The case of
Dicke-Brans-Jordan cosmologies was scrutinised by \citet{Rudiger82}
and \citet{Partovi84} studied the effects of inhomogeneities.
\citet{Seto06} suggested that a measurement of the anisotropy of
\zdot\ could be used to constrain the very low-frequency gravitational
wave background. Several of these authors have pointed out the
superior redshift accuracy achieved in the radio regime compared to
the optical, and \citet{Davis78} entertained the possibility of using
a $21$~cm absorption line at $z = 0.69$ in the radio spectrum of 3C
286 for a \zdot\ measurement. The detrimental effects of peculiar
velocities and accelerations, which may swamp the cosmic signal, were
discussed by \citet{Phillipps82}, \citet{Lake82} and \citet{Teuber86}.
Finally, \citet{Loeb98} first proposed the Lyman $\alpha$ (\lya)
forest as an appropriate target for a \zdot\ measurement
(acknowledging D.~Sasselov for the idea) and assessed the prospects
for a successful detection in the context of currently existing
observational technology. All except the last of these studies
concluded that a \zdot\ measurement was beyond the observational
capabilities of the time.

The purpose of the present paper is to examine the impact of the next
generation of $30$--$60$-m Extremely Large Telescopes (ELTs) on the
feasibility of determining \zdot. The key issue is obviously the
accuracy to which one can determine redshifts. In the absence of
systematic instrumental effects this accuracy depends only on the
intrinsic sharpness of the spectral features used, the number of
features available and the signal-to-noise ratio (S/N) at which they
are recorded (assuming that the features are resolved). In a
photon-noise limited experiment the latter in turn depends only on the
flux density of the source(s), the size of the telescope, the total
combined telescope/instrument efficiency and the integration time. In
this paper we will investigate this parameter space in detail,
expanding on previous work in several ways: in Section \ref{pottarg}
we confirm the usefulness of the \lya\ forest by quantifying the
peculiar motions of the absorbing gas using hydrodynamic simulations
of the intergalactic medium (IGM). In Section \ref{sims} we use Monte
Carlo simulations of the \lya\ forest to quantify how its properties
translate to a radial velocity accuracy, and we consider the
usefulness of other absorption lines in Section \ref{ext_spec}. These
results are then used in Section \ref{photons} where we explore the
observational parameter space and realistically assess the feasibility
of a \zdot\ experiment with an ELT. Finally, we summarise our findings
in Section \ref{conclusions}.

\section{Measuring the dynamics}
\label{dynamics}
In any metric theory of gravity one is led to a very specific form of
the metric by simply assuming that the Universe is homogeneous and
isotropic. The evolution in time of this so-called Robertson-Walker
metric is entirely specified by its global scale factor, $a(t)$. The
goal is to measure or reconstruct this function. Recall that we
observe the change of $a$ with time by its wavelength-stretching
effect on photons traversing the Universe. A photon emitted by some
object at comoving distance $\chi$ at time $\tem$ and observed by us
at $\tobs$ suffers a redshift of
\begin{equation}
\label{za}
1 + z(\tobs, \tem) = \frac{a(\tobs)}{a(\tem)}
\end{equation}
(where, obviously, only two of the three variables $\chi$, $\tem$ and
$\tobs$ are independent). If it were possible to measure not only a
photon's redshift but also its $\tem$, then the problem would be
solved by simply mapping out the present-day relation between redshift
and look-back time, i.e.\ $z(\tobs = t_0, \tem)$, where $t_0$ denotes
today.

A different approach to the problem is to consider how the redshift of
an object at a fixed comoving distance $\chi$ evolves with $\tobs$,
i.e.\ to consider the function $z_{\vert\chi}(\tobs)$. For a given
$\chi$, $\tobs$ determines $\tem$ and so we have dropped the
dependence on $\tem$. In principle, it is possible to map out
$z_{\vert\chi}(\tobs)$ (at least for $\tobs > t_0$), and this would be
the most direct determination of $a(t)$, no matter which object is
used. However, a full characterisation of $z_{\vert\chi}(\tobs)$ would
require observations over several Gyr. Over a much shorter timescale,
$\Delta \tobs$, one can at most hope to determine the first order term
of the Taylor expansion
\begin{equation}
\frac{\d z_{\vert\chi}}{\d \tobs}(\tobs) \approx
\frac{z_{\vert\chi}(\tobs + \Delta \tobs) -
z_{\vert\chi}(\tobs)}{\Delta \tobs}.
\end{equation}
As we will see presently, it turns out that measuring $\d
z_{\vert\chi}/\d \tobs$ is in fact sufficient to reach our goal
of reconstructing $a(t)$. By differentiating equation \eref{za} with
respect to $\tobs$ we find
\begin{equation}
\frac{\d z_{\vert\chi}}{\d \tobs}(\tobs) =
[1 + z_{\vert\chi}(\tobs)] H(\tobs) - H(\tem),
\end{equation}
where we have used that $\d \tobs = [1 + z(\tobs)] \, \d \tem$ for a
fixed $\chi$, and $H = \dot a a^{-1}$. Evaluating at $\tobs = t_0$,
replacing the unknown $\tem$ with its corresponding redshift, and
dropping the reminder that we are considering the redshift of an
object at a fixed distance $\chi$ we simply obtain \citep{McVittie62}:
\begin{equation}
\label{zdot}
\dot z \equiv \frac{\d z}{\d \tobs}(t_0) = (1+z) H_0 - H(z).
\end{equation}
\zdot\ is a small, systematic drift as a function of time in the
redshift of a cosmologically distant source as observed by us
today. This effect is induced by the de- or acceleration of the
expansion, i.e.\ by the change of the Hubble parameter $H$. Since
$H_0$ is known \citep{Freedman01}, this drift is a direct measure of
the expansion velocity at redshift $z$. Measuring \zdot\ for a
number of objects at different $z$ hence gives us the function $\dot
a(z)$. The point is that given $a(z)$ {\em and} $\dot a(z)$, one can
reconstruct $a(t)$. A measurement of $\dot z(z)$ therefore amounts to
a purely dynamical reconstruction of the expansion history of the
Universe.

\begin{figure}
\psfig{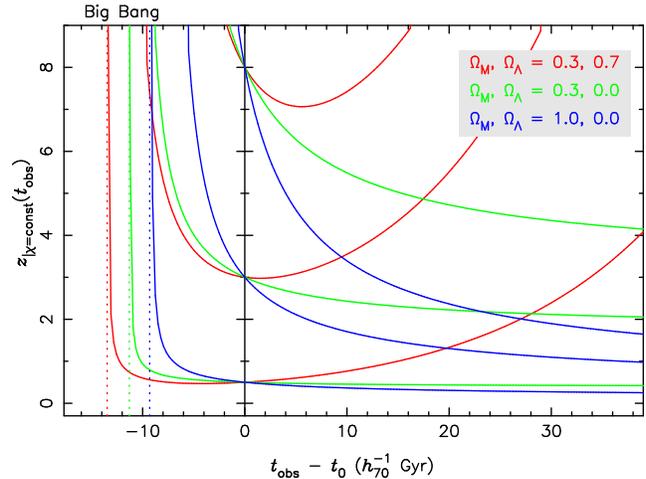}
\caption{Redshift evolution of three objects with present-day
  redshifts $z(t_0) = 0.5, 3$ and $8$ as a function of time of
  observation, for three different combinations of $\om$ and $\ol$ as
  indicated. For each case the dotted lines indicate the Big Bang.
  $\tobs$ is shown relative to the present day, $t_0$.}
\label{zt}
\end{figure}

Predicting the redshift drift $\dot z(z)$ requires a theory of
gravity. Inserting the Robertson-Walker metric into the theory's field
equation results in the Friedman equation, which specifically links
the expansion history with the densities, $\Omega_i$, and equation of
state parameters, $w_i$, of the various mass-energy components of the
Universe. In the case of general relativity the Friedman equation is
given by:
\begin{equation}
\label{fried}
H(z) = H_0 \left[ \sum_i \Omega_i (1+z)^{3(1+w_i)} + \ok (1+z)^2 
\right]^{\frac{1}{2}},
\end{equation}
where $\ok = 1 - \sum \Omega_i$. Here we consider only two components:
cold (dark) matter (CDM) with $w_{\rm M} = 0$ and dark energy in the
form of a cosmological constant with $w_\Lambda = -1$. In
Fig.~\ref{zt} we plot $z_{\vert\chi}(\tobs)$ for three different
objects [chosen to have $z(t_0) = 0.5, 3$ and $8$] and for three
different combinations of $\om$ and $\ol$, where we have also assumed
$H_0 = 70 \; h_{70}$\kms~Mpc$^{-1}$. For each object, the redshift
goes to infinity at some time in the past when the object first
entered our particle horizon. If $\ol = 0$ the redshift continually
decreases thereafter as the expansion is progressively slowed down by
$\om$. Hence, in this case the redshifts of all objects are decreasing
at the present time.

\begin{figure}
\psfig{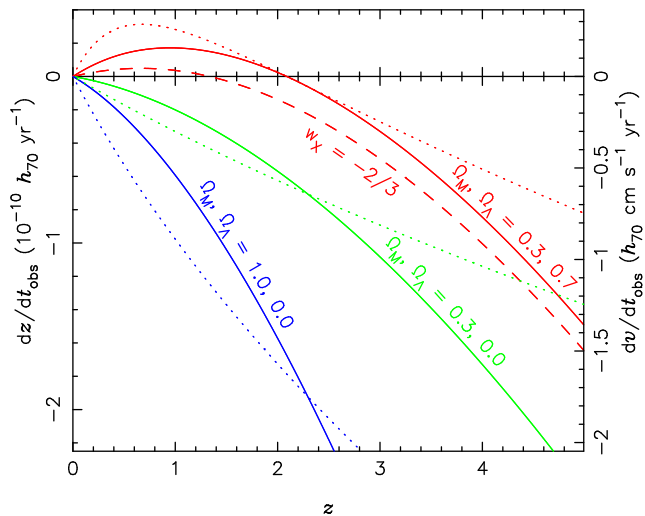}
\caption{The solid (dotted) lines and left (right) axis show the
  redshift drift \zdot\ ($\vdot$) as a function of redshift for
  various combinations of $\om$ and $\ol$ as indicated. The dashed
  line shows \zdot\ for the case of dark energy having a constant $w_X
  = -\frac{2}{3}$ (and $\om, \Omega_X = 0.3, 0.7$).}
\label{dzdt}
\end{figure}

However, if $\ol \neq 0$ the initial decrease is followed by a
subsequent rise due to $\Lambda$ relieving matter as the dominant
mass-energy component and causing the expansion to accelerate. The
turn-around point may lie either in the past or in the future,
depending on the object's distance from us; i.e.\ if $\ol \neq 0$ an
object's redshift may be either increasing or decreasing at the
present time. For distant (nearby) objects, the Universe was mostly
matter ($\Lambda$)-dominated during the interval $[\tem, t_0]$ and
hence underwent a net deceleration (acceleration), resulting in $\dot
z < (>) \; 0$ at the present time (see \citealp{Gudmundsson02} for
a more detailed discussion).

These features are evident in Fig.~\ref{dzdt}, where we plot the
expected present-day redshift drift, $\dot z(z)$, for various values
of $\om$ and $\ol$ (solid lines), and for a case where the dark energy
$w \neq -1$ (dashed line). The redshift drift is also shown in
velocity units (dotted lines), where $\vdot = c \, \dot z \,
(1+z)^{-1}$. As noted above, the existence of a redshift region where
$\dot z > 0$ is the hallmark of $\ol \neq 0$. Note also the scale of
Fig.~\ref{dzdt}. At $z = 4$ the redshift drift is of order $10^{-9}$
or $6$\cms\ per decade. For comparison, the long-term accuracy
achieved in extra-solar planet searches with the high-resolution
echelle spectrograph HARPS on the ESO 3.6-m Telescope is of order
$1$~m~s$^{-1}$ \citep[e.g.][]{Lovis05}.

\section{Choosing an accelerometer}
\label{pottarg}
A priori, it is not at all obvious which spectral features of which
set or class of objects might be best suited for a \zdot\
measurement. Clearly though, potential candidate targets should boast
as many of the following desirable attributes as possible. (i)~They
should faithfully trace the Hubble flow. Although peculiar motions are
expected to be random with respect to the Hubble flow, the additional
noise introduced by them could potentially conceal the cosmic signal
(\citealp{Phillipps82,Teuber86}; but see also Appendix \ref{vpec}).
(ii)~The targets should have the sharpest possible spectral features
to minimise the error on individual redshift measurements. (iii)~The
number of useful spectral features per target should be as high as
possible in order to maximise the amount of relevant information per
unit observing time. (iv)~The targets should be as bright as possible
and finally, (v)~they should exist over a wide redshift range, and
particularly at high $z$, where the signal is expected to be largest.

Clearly, several of these features are in conflict with each
other. Requirements (i) and (ii) are conflicting because sharp
spectral features require cold material which is generally found in
dense regions inside deep potential wells, which in turn generate
large peculiar accelerations. Similarly, point (i) clashes with point
(iv) because high intrinsic luminosities require very massive and/or
highly energetic processes, again implying deep potential wells. Since
very dense regions are relatively rare there is likewise tension
between requirements (ii) and (iii). Finally, the demands for
brightness and high redshift are also difficult to meet
simultaneously.

Hence, it seems that it is impossible to choose a class or set of
objects that is the ideal target, in the sense that it is superior to
every other class or set in each of the above categories. However,
there is one class of `objects' that meets all but one of the
criteria.

\subsection{The Lyman $\alpha$ forest}
\label{lyaf}
The term `\lya\ forest' refers to the plethora of absorption lines
observed in the spectra of all quasi-stellar objects (QSOs) shortwards
of the \lya\ emission line.\footnote{Since the \lya\ forest is only
observable from the ground for $z \ga 1.7$ we only consider the
high-$z$ \lya\ forest in this paper. We also exclude the higher column
density Ly limit and damped \lya\ (DLA) absorbers from the
discussion.} Almost all of this absorption arises in intervening
intergalactic \ion{H}{i} between us and the QSO (see \citealp{Rauch98}
for a review). Since the absorbing gas is physically unconnected with
the background source against which it is observed we elegantly avoid
the conflict between requirements (i) and (iv) above. However, as the
gas is in photoionization equilibrium with the intergalactic
ultraviolet (UV) background, its temperature is of order $10^4$~K
(\citealp*{Theuns00}; \citealp{Schaye00b}). Consequently, the
absorption lines are not particularly sharp and the typical line width
is $\sim$$30$\kms\ \citep*[e.g.][]{Kim01}. On the other hand, QSOs are
among the brightest sources in the Universe and exist at all redshifts
out to at least $\sim$$6$ \citep{Fan06}. Furthermore, each QSO
spectrum at $z \ga 2$ shows on the order of $10^2$ absorption
features.

In the following we will consider the question whether, apart from the
Hubble expansion, other evolutionary processes acting on the absorbing
gas might also significantly affect the measured positions of the
absorption lines over the timescale of a decade or so. First and
foremost is the issue to what extent the \lya\ forest is subject to
peculiar motions.

\subsubsection{Peculiar motions}
Cosmological hydrodynamic simulations (\citealp{Cen94};
\citealp*{Zhang95}; \citealp{Hernquist96}; \citealp{MiraldaEscude96};
\citealp{Cen97}; \citealp{Charlton97}; \citealp{Theuns98b};
\citealp{Zhang98}), analytic modelling \citep[e.g.][]{Bi97,Viel02} and
observations of the sizes and shapes of the absorbers
(\citealp{Bechtold94}; \citealp{Dinshaw94}; \citealp{Smette95};
\citealp*{Charlton95}; \citealp{DOdorico98}; \citealp{Liske00b};
\citealp{Rollinde03}; \citealp*{Becker04}; \citealp{Coppolani06}; see
also \citealp{Rauch95}) all suggest that the \lya\ forest absorption
occurs in the large-scale, filamentary or sheet-like structures that
form the cosmic web. These are at most mildly overdense and are
participating, at least in an average sense, in the general Hubble
expansion. However, at least some fraction of the absorbing structures
(depending on redshift) must be expected to have broken away from the
Hubble flow and to have begun collapsing under the influence of local
gravitational potential wells.

This exact issue was investigated by \citet{Rauch05} who studied the
distribution of velocity shear between high-redshift \lya\ absorption
common to adjacent lines of sight separated by $1$ to $300 \;
h_{70}^{-1}$~kpc. They showed that the observed shear distributions
were indeed in good agreement with the absorbing structures undergoing
large-scale motions dominated by the Hubble flow. In fact, the
distributions could be reproduced very convincingly by artificial
pairs of spectra created from a hydrodynamical simulation of the
IGM. This result instills us with further confidence that such
simulations accurately capture the kinematics of the gas responsible
for the \lya\ forest.

Thus assured, we will now explicitly examine the peculiar velocities
and accelerations of the absorbing gas in a hydrodynamic simulation of
the IGM produced by the parallel TreeSPH code {\sc gadget-2}
\citep{Springel05}. We have used this code in its TreePM mode in order
to speed up the calculation of long-range gravitational forces. The
simulations were performed with periodic boundary conditions and with
$400^3$ dark matter and $400^3$ gas particles in a box of $60 \;
h_{100}^{-1}$~Mpc comoving size. Radiative cooling and heating
processes were followed using an implementation similar to that of
\citet*{Katz96} for a primordial mix of hydrogen and helium. The UV
background was taken from \citet{Haardt96}. In order to further
increase the speed of the simulation we applied a simplified star
formation criterion: all of the gas at overdensities $> 10^3$ times
the mean density and with temperature $< 10^5$~K was turned into
stars. The cosmological parameters were set to $\om= 0.26$, $\ol =
0.74$, $\Omega_{\rm b} = 0.0463$, $n_s = 0.95$, $\sigma_8 = 0.85$ and
$H_0 = 72$\kms~Mpc$^{-1}$, where $\Omega_{\rm b}$ is the baryonic
density parameter, and $n_s$ and $\sigma_8$ are the spectral index and
amplitude of the linear dark matter power spectrum,
respectively. These values are in excellent agreement with recent
joint analyses of CMB, SNIa, galaxy clustering and \lya\ forest data
(\citealp{Spergel07}; \citealp{Seljak06}; \citealp*{Viel06}). The
\lcdm\ transfer function was computed with {\sc cmbfast}
\citep{Seljak96}. 
The above corresponds to the simulation series B2 of \citet*{Viel04}
which has been widely used for cosmological studies. Moreover, this is
the exact same simulation that successfully reproduced the velocity
shear distributions observed by \citet{Rauch05}.

\begin{figure}
\psfig{file=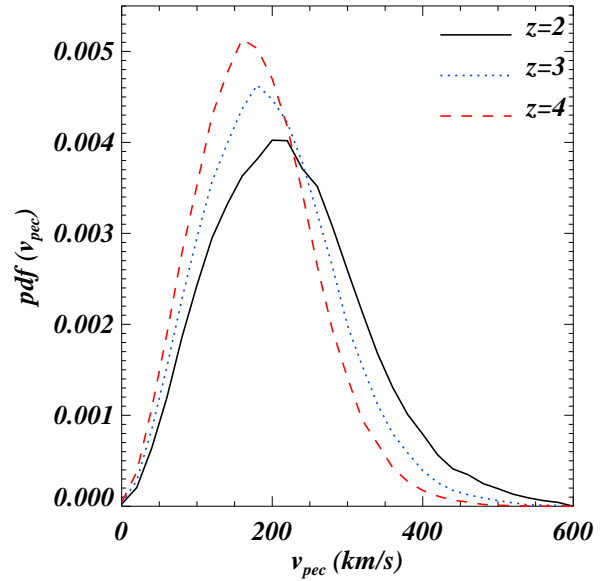,width=\columnwidth}
\caption{Probability distribution functions (PDFs) of the peculiar
  velocity of the absorbing gas along $1000$ random lines of sight
  through our simulation box at three different redshifts as
  indicated. Note that we are {\em not} using the value of the
  velocity's component along the LOS, but rather the modulus of the
  full 3-dimensional velocity.}
\label{vp}
\end{figure}

\begin{figure}
\psfig{file=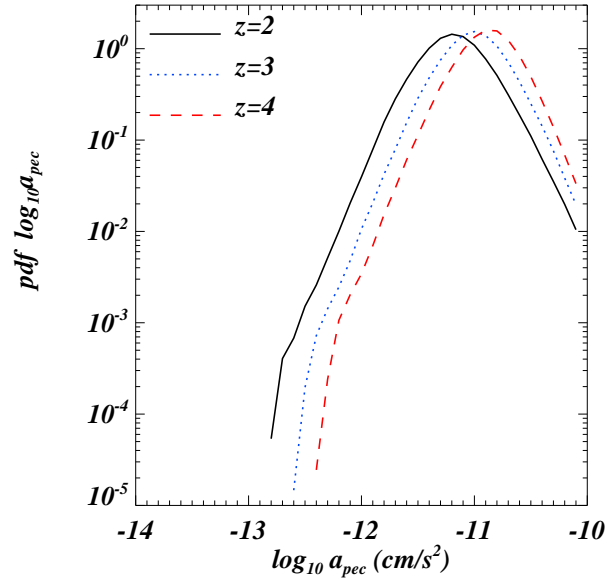,width=\columnwidth}
\caption{As Fig.~\ref{vp} for the peculiar acceleration.}
\label{ap}
\end{figure}

We pierced our simulation box with $1000$ random lines of sight (LOS)
and noted the physical properties of the absorbing gas along these
lines at three different redshifts ($z=2,3$ and $4$). In
Figs.~\ref{vp} and \ref{ap} we plot the distributions of peculiar
velocity and acceleration, respectively. The acceleration of the gas
is computed from its velocity by dividing the latter by a dynamical
time $t_{\rm dyn} = (G \rho)^{-1/2}$ \citep[e.g.][]{Schaye01}, where
$\rho$ is the total matter density. Evidently, the distributions do
not evolve rapidly with redshift. Since a given parcel of gas cannot
be decelerated, peculiar velocities cannot decrease with time and
hence their distribution shifts to slightly larger values at lower
redshifts. The average peculiar acceleration, on the other hand,
decreases with time because by volume the Universe is dominated by
low-density regions in which the gas density decreases as the Universe
expands.


Peculiar motions are expected to be randomly oriented with respect to
the LOS. Hence, when averaging over a large number of individual
\zdot\ measurements, peculiar motions will only introduce an
additional random noise component but no systematic bias. In Appendix
\ref{vpec} we explicitly derive an expression for the observed
redshift drift in the presence of peculiar motions. This expression
can be used to translate the distributions in Figs.~\ref{vp} and
\ref{ap} into the corresponding error distribution on \zdot. We find
that for a decade-long experiment the error due to peculiar motion is
of order $\sim$$10^{-3}$\cms. This must be compared to the error
induced by photon noise for an individual \zdot\ measurement from a
{\em single} absorption line. Clearly, if we wish to be able to detect
the redshift drift over a decade or so the overall accuracy of the
whole experiment has to be of order $\sim$$1$\cms. Since this will be
achieved using hundreds of absorption lines the error on an individual
\zdot\ measurement will be at least a factor of $\sim$$10$ larger. The
error due to peculiar motions is therefore sub-dominant by at least
$\sim$$4$ orders of magnitude. Hence we conclude that peculiar motions
will have no detrimental impact whatsoever on a redshift drift
experiment targeting the \lya\ forest.

\subsubsection{Galactic feedback}
\label{feedback}
So far we have only considered peculiar motions induced by gravity. In
principle, there are of course a number of non-gravitational ways of
imparting kinetic energy to the absorbing gas, mostly involving
galactic feedback. The main reason to believe that galactic feedback
must have had a far-reaching impact on the \lya\ forest is its early
and widespread low-level metal enrichment, even at fairly low column
densities (e.g.\
\citealp{Cowie95,Ellison00,Schaye00a,Schaye03,Aguirre04};
\citealp*{Simcoe04}). Although very low-level enrichment could be
achieved in situ with Population III stars \citep[e.g.][]{Gnedin97}
the general consensus is that some sort of mechanism is required to
transport metals from (proto-)galaxies into the IGM. In contrast, no
consensus has been reached as to which of the possible mechanisms (or
combination thereof) is the correct one. Candidates are mergers and
tidal interactions \citep[e.g.][]{Gnedin98}, ram pressure stripping,
radiation pressure on dust grains \citep[e.g.][]{Aguirre01a},
photo-evaporation during reionization \citep{Barkana99} and SN-driven
winds, either from low-mass, (pre-)galactic halos at $z \approx 10$
\citep[e.g.][]{Madau01} or from massive starbursting galaxies at $z <
5$ \citep[e.g.][]{Aguirre01c,Adelberger03}. From our point of view the
last possibility is the most worrying as it has the highest potential
of significantly altering the kinematic structure of the IGM at the
time of observation.

There exists persuasive evidence of the existence of galactic
superwinds from Lyman break galaxies at $z \approx 3$
\citep[e.g.][]{Pettini01} and it is likely that some fraction of
strong metal absorption lines are connected with these structures
\citep[e.g.][]{Adelberger03,Simcoe06}. On the other hand, there is no
observational evidence at all that superwinds are significantly
stirring up the high redshift \lya\ forest at the time we observe
it. Using column density and optical depth differences across a close
pair of lines of sight \citet{Rauch01} found no indication in the
forest's small-scale density structure for widespread recent
disturbances. Similarly, as discussed above, the velocity shear
between adjacent lines of sight is entirely explained by the Hubble
flow and gravitational instability \citep{Rauch05}, leaving little
room for non-gravitationally induced motion. Indeed, the simple fact
that the aforementioned hydrodynamical simulations -- which did {\em
not} include any galactic feedback -- were so successful in
reproducing the observed properties of the \lya\ forest, including its
line broadening distribution and clustering, raises the question of
how significant amounts of feedback could be integrated without
upsetting the existing agreement between the models and the data
\citep*{Theuns01}. It seems that the volume filling factor of galactic
superwinds is limited to a few per cent
\citep{Theuns02,Desjacques04,Pieri04,Cen05,Bertone05}. We conclude
that, whatever the process of metal enrichment may be, there is
currently no reason to believe that it has a wholesale effect on the
kinematics of the general IGM (as probed by the \lya\ forest) at the
time of observation.

\subsubsection{Optical depth variations}
\label{tauv}
Consider the gas responsible for a given \lya\ forest absorption
feature. If the physical properties of the gas change over the
timescale of a decade or so this will cause a variation of the
feature's optical depth profile. Potentially, this could in turn lead
to a small shift in the feature's measured position and hence mimic a
redshift drift. The precise magnitude of this additional \zdot\ error
component will depend on the method used to extract the signal but we
can gain an impression of the relevance of the effect by comparing the
expected optical depth variation to the apparent optical depth change
(at a fixed spectral position) induced by the redshift drift.

The gas properties we consider here are density, temperature and
ionization fraction. In the fluctuating Gunn Peterson approximation
\citep*[e.g.][]{Hui97b} the \lya\ optical depth is related to these
quantities by $\tau \propto (1 + \delta)^2 \, T^{-0.7} \,
\Gamma^{-1}$, where $\delta$ and $T$ are the gas overdensity and
temperature, respectively, and $\Gamma$ is the photoionization
rate. The interplay between photoionization heating of the gas and
adiabatic cooling leads to a tight relation between temperature and
density, which can be well approximated by $T = T_0 (1 +
\delta)^\gamma$ \citep{Hui97a}. Hence we obtain
\begin{equation}
\label{tau}
\tau \propto (1 + \delta)^{2-0.7\gamma} \, T_0^{-0.7} \, \Gamma^{-1}.
\end{equation}
How do these quantities evolve with time? According to linear theory
density perturbations grow as $(1+z)^{-1}$ . Although exactly true
only for an Einstein-de Sitter Universe, this represents an upper
limit in open and flat models with a cosmological constant. Hence we
will err on the side of caution by adopting this growth factor in the
following. The evolution of the temperature-density relation can be
gleaned from figure 6 of \citet{Schaye00b}: $\d \gamma/\d z = -0.1$
and $\d T_0 / \d z = 0.25 \, T_0$ (where we have conservatively used
the steep evolution between redshifts $2$ and $3$). We take the
evolution of $\Gamma$ from figure 7 of \citet{Bolton05}: $\d \Gamma /
\d z = -0.3 \Gamma$. Taking the appropriate derivatives of equation
\eref{tau}, inserting the above values and translating a decade in our
reference frame to a redshift difference $\d z$ at $z = 3$ we find an
optical depth variation of $\d \tau = 3 \times 10^{-9} f \tau$, where
$f \approx 0.2, 0.2, 0.05, 0.3$ for the variation of $\delta$, $T_0$,
$\gamma$ and $\Gamma$, respectively. As we will see in Section
\ref{genspec}, these values for $\d \tau$ are at least $\sim$$2$
orders of magnitude smaller than the typical optical depth changes due
to the redshift drift. We therefore conclude that changes in the
physical properties of the absorbing gas are not expected to interfere
with a \zdot\ measurement from the \lya\ forest.

\subsection{Molecular absorption lines}
\label{molecs}
Before we move on to investigate the details of a redshift drift
experiment using ELT observations of the \lya\ forest, let us briefly
digress here to consider a very different \zdot\ experiment using
another future facility. We have explored in some detail the
possibility of using the Atacama Large Millimeter Array (ALMA) to
measure \zdot\ from rotational molecular transitions seen in
absorption against background continuum sources. Going to the (sub-)mm
regime has the advantages of potentially very high resolution and less
photon noise for a given energy flux. Furthermore, molecular
absorption lines can be very sharp, with line widths as low as $\la
1$\kms. However, the molecular gas in nearby galaxies is strongly
concentrated towards the central regions. Hence we must expect the gas
to be subject to peculiar accelerations similar in magnitude to the
cosmological signal. That in itself would not necessarily be
problematic as long as we had many individual \zdot\ measurements from
different objects over which to average. Unfortunately, it seems
unlikely that this will be the case. At present, rotational molecular
lines have been detected in only four absorption systems \citep[with
redshifts $0.25$--$0.89$;][and references therein]{Wiklind99}, despite
intensive searches \citep[e.g.][]{Curran04}. Based on the incidence of
these lines and the number of continuum sources with fluxes larger
than $10$~mJy at $90$~GHz we estimate that the number of molecular
absorption systems observable with ALMA will be $\sim$$50$, with only
$5$--$10$ of these showing narrow lines -- not enough to overcome the
uncertainties due to peculiar motions. Hence we have decided not to
pursue the case for molecular absorption lines any further.

\section{Sensitivity of the Ly$\alpha$ forest to radial velocity shifts}
\label{sims}
In Section \ref{dynamics} we have seen that the redshift drift is a
very small effect. In order to detect it, an experiment must achieve
an overall accuracy with which radial velocity shifts can be
determined of order $\sim$$1$\cms. In this section we will investigate
how the properties of the \lya\ forest translate to a radial velocity
accuracy, $\sv$, and how $\sv$ depends on the instrumental
characteristics of the spectra and on redshift. Specifically, we would
like to know how many \lya\ forest spectra of which resolution and S/N
are needed at a given redshift to achieve the required $\sv$.

We will investigate these issues using artificial spectra.
High-resolution observations have demonstrated that, to first
approximation, the \lya\ forest can be decomposed into a collection of
individual absorption lines \citep[e.g.][]{Kim01}. These are usually
taken to be Voigt profiles and so each line is characterised by three
parameters: redshift, $z$, \ion{H}{i} column density, $N_{\rm HI}$,
and velocity width, $b$. Here, we will reverse this decomposition
process and generate (normalised) spectra with the desired
instrumental characteristics from given lists of absorption lines. We
will use two types of line lists. First, we will generate line lists
from Monte Carlo (MC) simulations based on the statistics of the
largest available samples of absorption lines. Secondly, to validate
our simulations, we will use 8 line lists available in the literature
that have previously been derived from high-resolution observations.

\subsection{Simulated absorption line lists}
\label{slists}
We form simulated MC line lists by simply randomly drawing values for the
absorption line parameters from their observed distributions
\cite[e.g.][]{Hu95,Lu96,Kirkman97,Kim97,Kim01,Kim02}:
\begin{equation}
\label{vp_distribution}
f(z, N_{\rm HI}, b) \; \propto \; (1+z)^\gamma \; N_{\rm HI}^{-\beta} \;
\exp \left[-\frac{(b - \bar b)^2}{2 \sigma_b^2}\right],
\end{equation}
where $\gamma = 2.2$, $\beta = 1.5$, $\bar b = 30$\kms\ and $\sigma_b
= 8$\kms. We impose limits of $15 < b < 100$\kms\ and also restrict
$N_{\rm HI}$ to the classical \lya\ forest regime, excluding Lyman
limit and DLA systems: $12 < \log N_{\rm HI} ($cm$^{-2}) < 16$. The
above distribution is normalised to give $10^2$ lines with $13.64 <
\log N_{\rm HI} ($cm$^{-2}) < 16$ per unit redshift at $z = 2$
\citep{Kim01}. The actual number of absorption lines in a given line
list is drawn from a Poisson distribution with a mean determined by
the normalisation.

The above MC approach allows us to quickly generate large amounts of
spectra with realistic characteristics. Note, however, that we make no
assumptions regarding the underlying physics of the IGM in which the
absorption occurs. We simply use the observational fact that \lya\
forest spectra can be well represented as a random collection of Voigt
profiles.

The most significant difference between our MC line lists and the real
\lya\ forest is clustering: the real \lya\ forest is not randomly
distributed in redshift but shows significant redshift-space
correlations on scales of at least $100$\kms\
\citep[e.g.][]{Cristiani95,FernandezSoto96,Liske00b}. The impact of
clustering will be discussed in detail in Section \ref{sigv_res}.

\subsection{Real absorption line lists}
\label{rlists}

\begin{table}
\begin{center}
\begin{minipage}{7.8cm}
\caption{Observed \lya\ forest line lists from the literature.}
\label{rqsos}
\begin{tabular}{lcccc}
\hline
QSO & $\zqso$ & $\lambda$ range$^a$ & $N_{{\rm Ly}\alpha}^b$ & Reference\\
\hline 
Q1101$-$264    & $2.145$ & $3226$$-$$3810$ & $290$ & 1\\
J2233$-$606    & $2.238$ & $3400$$-$$3850$ & $226$ & 2\\
HE1122$-$1648  & $2.400$ & $3500$$-$$4091$ & $354$ & 1\\
HE2217$-$2818  & $2.413$ & $3550$$-$$4050$ & $262$ & 2\\
HE1347$-$2457  & $2.617$ & $3760$$-$$4335$ & $362$ & 1\\
Q0302$-$003    & $3.281$ & $4808$$-$$5150$ & $223$ & 1\\
Q0055$-$269    & $3.655$ & $4852$$-$$5598$ & $535$ & 1\\
Q0000$-$26     & $4.127$ & $5380$$-$$6242$ & $431$ & 3\\
\hline
\end{tabular}\\
References: 1 = \citet{Kim02}, 2 = \citet{Kim01}, 3 = \citet{Lu96}.\\
$^a$Wavelength range covered by the \lya\ forest line lists in \AA.\\
$^b$Number of \lya\ forest lines.
\end{minipage}
\end{center}
\end{table}

We have collected 8 QSO absorption line lists from the literature (see
Table~\ref{rqsos}). These were derived from UVES/VLT (7 objects) or
HIRES/Keck data (1 object). All spectra have a resolution of
$\mbox{FWHM} \approx 7$~\kms, while the typical S/N per pixel varies
from $\sim$$10$ for Q0000$-$26 to $\sim$$50$ in the case of
HE2217$-$2818. In all cases the absorption line lists were derived by
fitting Voigt profiles to the spectra using VPFIT.\footnote{by
R.F.~Carswell et al., see
http://www.ast.cam.ac.uk/$\sim$rfc/vpfit.html.} Details of the data
acquisition and reduction, as well as of the line fitting and
identification processes are given by \citet{Kim01,Kim02} and
\citet{Lu96}.

We note that two of the spectra, those of Q1101$-$264 and Q0000$-$26,
contain DLAs, which effectively block out parts of the
spectra. However, this is only expected to have a significant effect
at high redshift, where the \lya\ forest line density is high, and so
we have excluded the affected spectral region only in the case of
Q0000$-$26.

\subsection{The second epoch}
Simulating a \zdot\ measurement requires a second epoch
`observation' of the same line of sight: we generate a second epoch
absorption line list from the original line list (both real and
simulated) by simply shifting the redshift of each line according to a
given cosmological model:
\begin{equation}
\Delta z_i = \dot z(z_i; H_0, \om, \ol) \; \Delta t_0,
\end{equation}
where $\Delta t_0$ is the assumed time interval between the first and
second epochs. Unless stated otherwise, we use the standard, general
relativistic cosmological model, assuming fiducial parameter values of
$H_0 = 70 \; h_{70}$\kms~Mpc$^{-1}$, $\om = 0.3$ and $\ol = 0.7$.

\subsection{Generating spectra}
\label{genspec}
Given an absorption line list a normalised spectrum is generated by
\begin{equation}
S(\lambda) = \exp \left\lbrace-\sum_i^{N_{\rm al}} 
\tau[\lambda_\alpha(1+z_i), N_{{\rm HI},i}, b_i] \right\rbrace,
\end{equation}
where $\lambda$ is the observed wavelength, $N_{\rm al}$ is the number
of absorption lines in the spectrum, $\tau$ is the optical depth of a
Voigt profile and $\lambda_\alpha = 1215.67$~\AA\ is the rest
wavelength of the \ion{H}{i} \lya\ transition. This spectrum is then
pixelised, using a pixel size of $0.0125$~\AA, and convolved with a
Gaussian line-spread function. Unless stated otherwise, we will use a
resolution element four times the pixel size, corresponding to a
resolution of $R = 100\,000$ at $5000$~\AA. We then add random noise
to the spectrum assuming Poisson statistics, i.e.\ we assume a purely
photon-noise limited experiment. All S/N values quoted in this paper
refer to the S/N per pixel in the continuum. To begin with, we will
only consider the spectral range between the assumed background QSO's
\lya\ and \lyb\ emission lines, i.e.\ the classical \lya\ forest
region. By construction this region cannot contain any \ion{H}{i}
transitions that are of higher order than \lya, and it covers an
absorption redshift range of $\lambda_\beta / \lambda_\alpha (1 +
\zqso) - 1 < z < \zqso$, where $\zqso$ is the QSO's redshift and
$\lambda_\beta = 1025.72$~\AA\ is the rest wavelength of the
\ion{H}{i} \lyb\ transition. In Sections \ref{sv_lyb} and \ref{sv_m}
we will also consider other spectral regions.

\begin{figure}
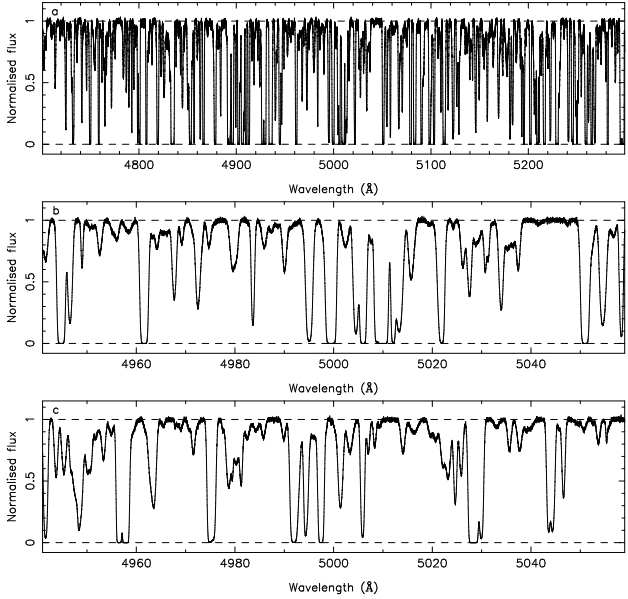

\psfig{file=codex_fig5a.ps,angle=-90,width=\columnwidth}
\psfig{file=codex_fig5b.ps,angle=-90,width=\columnwidth}
\psfig{file=codex_fig5c.ps,angle=-90,width=\columnwidth}
\caption{(a) Example of an artificial \lya\ forest spectrum at $z
  \approx 3$ generated from a Monte Carlo absorption line list. (b)
  Close-up of the region around $5000$~\AA. (c) Artificial spectrum
  generated from the observed line list of Q0302$-$003. Both spectra
  have $\mbox{S/N} = 100$.}
\label{sim_spec}
\end{figure}

\begin{figure}
\psfig{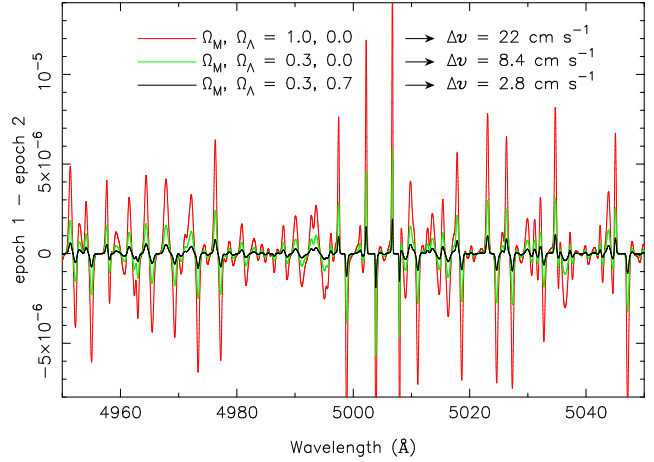}
\caption{Flux difference between two artificial, noiseless spectra of
  the same \lya\ forest at $z \approx 3$ simulated for two observing
  epochs separated by $\Delta t_0 = 10$~yr and for various combinations
  of $\om$ and $\ol$ as indicated. The redshift drift implied by these
  parameters is also given.}
\label{flux_diff}
\end{figure}

In Fig.~\ref{sim_spec} we show examples of artificial spectra
generated from both simulated (panels a and b) and real line lists
(panel c). The expected flux difference between two spectra of the
same absorption lines, taken a decade apart, is shown in
Fig.~\ref{flux_diff}. A similar plot for the optical depth difference
reveals that the variations discussed in Section \ref{tauv} are
unproblematic.

\subsection{Defining $\sv$}
\label{sigv}
In principle, a \zdot\ determination will involve the measurement of
radial velocity differences between the corresponding features of a
pair of spectra of the same object taken several years apart. We now
need a method to estimate the accuracy to which these differences can
be determined. This requires knowledge of how exactly the measurement
will be performed. However, a priori it is not at all obvious what the
optimal signal extraction method might be. To proceed nevertheless we
choose to base our analysis on the generic concept of the total radial
velocity information content of a spectrum, which was developed by
\citet*{Bouchy01} in the context of optimising radial velocity
searches for extra-solar planets.

Following \citet{Bouchy01} we begin by expressing the flux observed in
pixel $i$ at the second epoch as a small perturbation on the first
epoch flux in the same pixel:
\begin{equation}
\label{dvi}
S_{2i} = S_{1i} + \frac{\d S_i}{\d \lambda} \; \frac{\Delta v_i}{c} \;
\lambda_i,
\end{equation}
which defines a small velocity shift $\Delta v_i$ for each pixel.
$\lambda_i$ is the observed wavelength of the $i$th pixel and $\d
S_i/\d \lambda$ is the spectral slope of the flux at that pixel. To
first order the slope does not change between the two epochs and hence
it carries no epoch designation. Averaging the velocity shift over all
pixels in a spectrum, using weights $w_i$, we have:
\begin{equation}
\label{deltav}
\Delta v = \frac{\sum_i \Delta v_i \; w_i}{\sum_i w_i}.
\end{equation}
Clearly, the weight for the $i$th pixel should be chosen as the
inverse variance of $\Delta v_i$. In calculating this variance we must
differ from \citet{Bouchy01}. In the case of stars one of the spectra
can be assumed to be a perfect, noiseless template, essentially
because additional information on the same type of star can be used to
define it. However, since every \lya\ forest spectrum is unique we
cannot make the same assumption here, so that in our case both spectra
have noise. Hence we find:
\begin{equation}
\label{sigvi}
\sigma^2_{v_i} = \left (\frac{c}{\lambda_i \frac{\d S_i}{\d
      \lambda}}\right)^2 \left[ \sigma_{1i}^2 + \sigma_{2i}^2 \; +
      \; \frac{(S_{2i} - S_{1i})^2}{\left(\frac{\d S_i}{\d
      \lambda}\right)^2} \; \sigma_{S^\prime_i}^2\right],
\end{equation}
where $\sigma_{1i}$ and $\sigma_{2i}$ are the flux errors in the $i$th
pixel of the first and second epoch spectra, respectively, and
$\sigma_{S^\prime_i}$ is the error on the slope of the flux at pixel $i$.
We can see that a low weight is assigned to noisy pixels and to those
that have a small gradient, i.e.\ pixels in the continuum or in the
troughs of saturated absorption lines. Finally, with the above choice
of weights the error on $\Delta v$ is given by:
\begin{equation}
\label{sve}
\sv^2 = \frac{\sum_i \sigma^2_{v_i} w_i^2}{\left(\sum_i w_i\right)^2} 
= \frac{1}{\sum_i \sigma^{-2}_{v_i}}.
\end{equation}

The above process has the advantage of conveniently attaching a single
figure of merit to a given pair of spectra in a non-parametric,
model-independent way: $\sv$ simply represents the fundamental
photon-noise limit of the accuracy to which an overall velocity shift
between the two spectra can be determined. It is essentially just a
measure of the `wiggliness' and of the S/N of the spectra. However, we
point out that $\sv$ does not entirely capture all of the information
contained in a pair of spectra with respect to a \zdot\
measurement. From Fig.~\ref{dzdt} it is clear that the difference
between the first and second epoch spectra is not simply an overall
velocity shift. The second epoch spectrum will also be slightly
compressed with respect to the first epoch spectrum because the
redshift drift is larger at higher redshifts (i.e.\ longer
wavelengths) than at lower redshifts.  The additional information that
is contained in this small alteration of the spectrum's shape is
entirely ignored by the above method [because of the simple averaging
operation in equation \eref{deltav}] and hence it is clearly a
sub-optimal method of estimating the sensitivity of a pair of spectra
to \zdot. Despite this shortcoming we will retain the above definition
of $\sv$ for the sake of its simplicity.

\subsection{Results}
\label{sigv_res}
We are now ready to derive the relevant scaling relations for $\sv$.
We begin by making two points regarding equation \eref{sigvi}. First,
we note that for a fixed total integration time (the sum of the
integration times spent observing the first and second epoch spectra)
the sum $\sigma_{1i}^2 + \sigma_{2i}^2$ in equation \eref{sigvi} takes
on its minimum when the first and second epoch integration times are
equal. Hence, the smallest possible $\sv$ is only achieved when the
spectra of both epochs have the same S/N. In the following we will
assume that this is the case. Secondly, from equation \eref{sigvi} it
is clear that $\sv$ scales as (S/N)$^{-1}$, as expected for a
photon-noise limited experiment.

Consider now a set of $\nqso$ targets that all lie at the same
redshift, each of which has been observed at two epochs such that all
$2 \nqso$ spectra have the same S/N. Again, since we are considering a
purely photon-noise limited experiment, $\sv$ should scale as $N_{\rm
pix}^{-1/2}$ (where $N_{\rm pix}$ is the total number of independent
data points in the sample) and hence also as
$\nqso^{-1/2}$. Furthermore, in this ideal case, it is irrelevant how
the {\em total} S/N is divided among the targets. Hence, for
simplicity we will use $\nqso = 1$ in the following.

\begin{figure}
\psfig{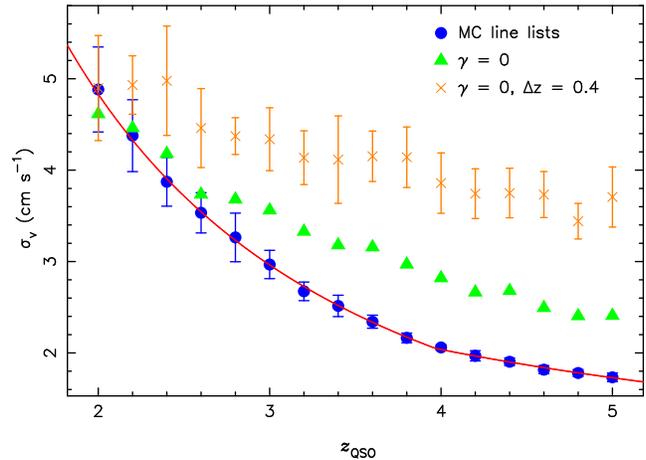}
\caption{The blue dots with error bars show the accuracy with which a
  radial velocity shift can be determined from a \lya\ forest spectrum
  as a function of QSO redshift. The red line parameterises the
  redshift dependence as in equation (\ref{svze}). Each point is the
  mean $\sv$ measured from $10$ pairs of artificial spectra with
  $\mbox{S/N} = 13\,000$, generated from simulated MC absorption line
  lists. The error bars show the $\pm1$ r.m.s.\ range of the $10$
  simulations. The green triangles show the results for simulations
  where the redshift evolution of the \lya\ forest has been switched
  off, i.e.\ where $\gamma = 0$ (cf.\ equation
  \ref{vp_distribution}). The orange crosses show the result of also
  restricting the $\sv$ measurement in each spectrum to a redshift
  path of constant length $\Delta z = 0.4$, as opposed to using the
  full \lya\ forest region between the QSO's \lya\ and \lyb\ emission
  lines.}
\label{sv_z}
\end{figure}

We now examine the behaviour of $\sv$ as a function of redshift using
the MC absorption line lists. For various QSO redshifts in the range
$2 \le \zqso \le 5$ we have generated $10$ pairs of \lya\ forest line
lists and spectra as described in Sections \ref{slists} to
\ref{genspec}, where each spectrum was given a S/N of $13\,000$. We
then measured each pair's $\sv$ according to equation (\ref{sve}). The
result is shown as blue dots in Fig.~\ref{sv_z}, where each point and
error bar is the mean and $\pm 1$ r.m.s.\ of the $10$ individual
measurements at each redshift. We stress that we are plotting the
expected accuracy of a velocity shift measurement performed on a
single pair of spectra of a single target at a given redshift, where
each spectrum has $\mbox{S/N} = 13\,000$. We are {\em not} plotting
the {\em combined} accuracy of $10$ such pairs.

From Fig.~\ref{sv_z} we can see that the radial velocity sensitivity
improves rather rapidly with redshift for $\zqso < 4$, but the
decrease is somewhat shallower at $\zqso > 4$. Overall, $\sv$ improves
by a factor of almost $3$ when moving from $\zqso = 2$ to
$5$. Specifically, we find:
\begin{equation}
\label{svze}
\sv \propto \left\lbrace
\begin{array}{ll}
(1 + \zqso)^{-1.7} \qquad & \zqso < 4 \\
(1 + \zqso)^{-0.9}  & \zqso > 4
\end{array} \right.
\end{equation}
The behaviour of $\sv$ as a function of redshift is due to the
combination of several factors. The first is the redshift evolution of
the \lya\ forest line density (cf.\ equation \ref{vp_distribution}).
At higher redshift more spectral features are available for
determining a velocity shift and so $\sv$ decreases. However, from $z
\approx 4$ the absorption lines severely blanket each other and the
number of sharp spectral features does not increase as rapidly
anymore, causing the flattening of $\sv$ at $\zqso > 4$. The green
triangles in Fig.~\ref{sv_z} show the $\sv$ measurements that result
from simulations where the redshift evolution of the \lya\ forest has
been switched off, i.e.\ where the evolutionary index $\gamma$ has
been set to $0$ (cf.\ equation \ref{vp_distribution}). Indeed, we can
see that in this case there is no evidence of a break.

Secondly, we recall that each \lya\ forest spectrum covers the entire
region between the QSO's \lya\ and \lyb\ emission lines. The redshift
path length of this region is given by $\Delta z = 0.156 (1 + \zqso)$.
Hence the number of independent pixels per spectrum also increases as
$(1 + \zqso)$. Since the S/N per pixel is kept constant this implies a
larger number of photons per spectrum and hence an improved
sensitivity to radial velocity shifts. The effect of this can be seen
by comparing the green triangles in Fig.~\ref{sv_z} with the orange
crosses which are the result of using a constant redshift path length
of $\Delta z = 0.4$ for each $\sv$ measurement, as well as $\gamma =
0$.

Finally, with $\gamma = 0$ and $\Delta z =$ const, the sensitivity to
{\em wavelength} shifts should be constant as a function of $\zqso$,
and so the sensitivity to {\em velocity} shifts should go as $(1 +
\zqso)^{-1}$ (as can be seen from equation \ref{sigvi}). In fact, the
orange crosses in Fig.~\ref{sv_z} decrease more slowly than this
because the widths of the absorption lines in wavelength space
increase as $(1 + \zqso)$, making the edges of the lines less steep
and hence slightly decreasing the spectrum's sensitivity to wavelength
shifts.

\begin{figure}
\psfig{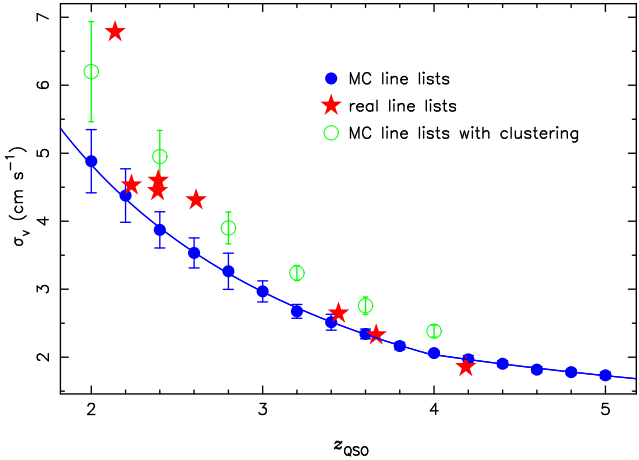}
\caption{Comparison of $\sv$ measurements derived from simulated MC
  line lists (blue dots and solid line, same as in Fig.~\ref{sv_z})
  and from real line lists (red stars). The green circles show the
  results derived from simulated MC line lists that include a simple
  scheme for clustering absorption lines in redshift space (see text
  for details).}
\label{svr_z}
\end{figure}

In Fig.~\ref{svr_z} we compare these results derived from the MC line
lists to those from the real line lists. From Table~\ref{rqsos} we can
see that the real line lists (and hence the corresponding spectra) do
not cover the full \lya\ forest regions, with differently sized pieces
missing both at the low and high redshift ends. Therefore we must
correct the $\sv$ values derived from the real line lists in order to
make them directly comparable to the values from the simulated
lists. The correction is achieved by first assigning a new, slightly
different QSO redshift to each spectrum, such that the `missing' low
and high redshift parts of the \lya\ forest region are equally
large. We then decrease the measured $\sv$ by a factor $(\Delta z_{\rm
obs} / \Delta z)^{1/2}$, where $\Delta z_{\rm obs}$ is the redshift
path length covered by the observed line list and $\Delta z$ is the
redshift path length of the full \lya\ forest region at the new
$\zqso$. The correction factors range from $0.56$ to $0.99$.

The red stars in Fig.~\ref{svr_z} show the corrected $\sv$ values
derived from single pairs of spectra generated from the real
absorption line lists with $\mbox{S/N} = 13\,000$, while the blue dots
show the measurements from the MC line lists (same as in
Fig.~\ref{sv_z}).  Overall the agreement between the results from the
MC and real line lists is very good, particularly at high redshift. At
$\zqso \approx 2.4$ the $\sv$ values from the real line lists are
$\sim$$15$ per cent higher than those from the MC lists. By far the
most significant deviation occurs at the lowest redshift where the
$\sv$ of Q1101$-$264 is higher than expected by $47$ per
cent. However, this is not too surprising as the line of sight towards
Q1101$-$264 is known to pass through an unusually low number of
absorbers with $\log N_{\rm HI}($cm$^{-2}) > 14$ \citep{Kim02}.

We believe that the small differences at $\zqso \approx 2.4$ are
mainly due to clustering of real absorption lines in redshift
space. Clustering has the effect of reducing the number of spectral
features because it increases line blanketing. However, at high
redshift line blanketing is already severe because of the high line
density and so clustering has a relatively smaller effect at high
redshift than at low redshift. We demonstrate that clustering can
explain the observed differences by generating a new set of MC line
lists which incorporate a toy clustering scheme: first, we randomly
draw the positions of `cluster' centres from the \lya\ forest redshift
distribution. We then populate each `cluster' with $n \ge 0$
absorbers, where $n$ is drawn from a Borel distribution
\citep{Saslaw89}. Since $n$ can be $0$, $1$ or $> 1$ this process
generates voids, single `field' absorbers, as well as groups and
clusters of lines. Finally, absorption lines are distributed around
their `host' clusters according to a Gaussian distribution with
$\mbox{FWHM} = 120$~\kms. The $\sv$ values that result from this new
set of MC line lists are shown as open green circles in
Fig.~\ref{svr_z}. The increase of $\sv$ compared to the unclustered
simulations is clearly very similar to that observed for the real line
lists and we conclude that clustering can indeed explain the small
difference between the results obtained from the observed and
simulated line lists at $\zqso \approx 2.4$. In any case, the near
coincidence of the $\sv$ value of J2233$-$606 at $\zqso = 2.23$ with
the expected value demonstrates that not all lines of sight are
adversely affected by clustering. In the following we will assume that
such sightlines can be pre-selected and hence we will ignore the
effects of clustering in the rest of this section.

\begin{figure}
\psfig{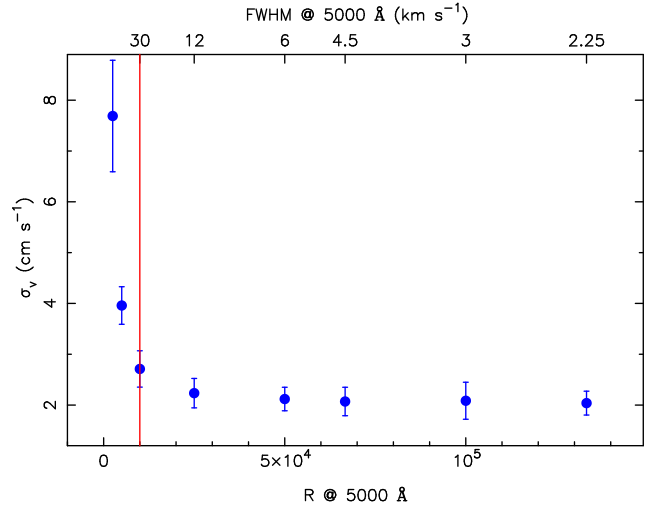}
\caption{Radial velocity accuracy as a function of spectral
  resolution. Each point and error bar is the mean and $\pm 1$ r.m.s.\
  of $10$ simulations at $\zqso = 4$ and $\mbox{S/N} = 13\,000$. The
  pixel size is kept constant and is chosen such that a resolution
  element is sampled by $3$ pixels at the highest resolution
  considered. The labels along the bottom axis denote the resolving
  power at $5000$~\AA, while the labels along the top axis show the
  equivalent FWHM of an unresolved line. The vertical red line marks
  the mode of the absorption lines' $b$ parameter distribution (cf.\
  equation \ref{vp_distribution}).}
\label{sv_fwhm}
\end{figure}

We now turn to the behaviour of $\sv$ as a function of spectral
resolution. For various resolving powers in the range $2500 \le R \le
1.33 \times 10^5$ we have generated $10$ pairs of line lists and
spectra with $\zqso = 4$ and $\mbox{S/N} = 13\,000$, and measured
their $\sv$ values as before. The result is presented in
Fig.~\ref{sv_fwhm}, where we show the resolving power along the bottom
axis and the corresponding FWHM of an unresolved line along the top
axis. We stress that the pixel size was the same for all spectra ($=
0.0125$~\AA) and that it was chosen such that a resolution element was
well sampled even at the highest resolution considered. Hence the
strong increase of $\sv$ towards lower $R$ in Fig.~\ref{sv_fwhm} is
not due to different numbers of photons per spectrum or sampling
issues. Instead, it is simply due to the loss of information caused by
convolving the spectrum with a line-spread function that is broader
than the typical intrinsic absorption line width (marked by the
vertical line in Fig.~\ref{sv_fwhm}). Indeed, at $R \ga 30\,000$ the
\lya\ forest is fully resolved and in this regime $\sv$ is independent
of $R$.

Summarising all of the above we find that the accuracy with which a
radial velocity shift can be determined from the \lya\ forest
scales as:
\begin{equation}
\label{sveq}
\sv = 2 \left(\frac{\mbox{S/N}}{2370}\right)^{-1} 
\left(\frac{\nqso}{30}\right)^{-\frac{1}{2}}
\left(\frac{1 + \zqso}{5}\right)^{-1.7} \mbox{\cms},
\end{equation}
where the last exponent changes to $-0.9$ at $\zqso > 4$, and where the
same S/N (per pixel) is assumed for all $\nqso$ spectra at both epochs.

\section{Sensitivity gains from other spectral regions}
\label{ext_spec}
In the previous section we have investigated the sensitivity of the
\lya\ forest to radial velocity shifts, using only the \ion{H}{i}
\lya\ transition in the region between a QSO's \lya\ and \lyb\
emission lines. However, modern echelle spectrographs are capable of
covering a much wider spectral range in a single exposure and so the
question arises whether other spectral regions, containing absorption
lines from other ions or other \ion{H}{i} transitions, can expediently
contribute towards a \zdot\ measurement.

Before applying the procedure of the previous section to more extended
spectra, we briefly follow-up on the discussion at the end of Section
\ref{sigv}, where we pointed out that $\sv$ does not capture the
information that is contained in the \zdot-induced change of the shape
of a spectrum. In the case of a \lya-only spectrum this change of
shape consisted of a compression of the spectrum. If we allow
additional transitions with different rest wavelengths then we no
longer have a one-to-one correspondence between absorption redshift
and wavelength, and so the redshift drift will in general induce a
much more complex change of the shape of a spectrum. Hence it is clear
that any attempt to harness this additional information must involve
the complete identification and modelling of all absorption features
used in the analysis.

Indeed, the complete identification of all metal absorption lines will
be necessary in any case, even if one endeavours to measure the
redshift drift only from the \lya\ forest. The point is that the \lya\
forest may of course be `contaminated' by metal lines. Since these
lines may arise in absorption systems that lie at completely different
redshifts compared to the \lya\ lines, their redshift drift may also
be very different. Hence, any lines wrongly assumed to be \lya\ could
potentially lead to erroneous results. The only way to avoid such
biases is to completely identify all absorption features.

\subsection{The Ly$\beta$ forest}
\label{sv_lyb}
Each \ion{H}{i} absorption line in the \lya\ forest has corresponding
counterparts at shorter wavelengths that result from the higher order
transitions of the Lyman series. Obviously, all the arguments
concerning the suitability of the \lya\ forest for a \zdot\
measurement also apply to these higher order lines, and so, in
principle, they should be almost as useful as the \lya\ lines (where
the qualifier `almost' is owed to the decreasing optical depth with
increasing order). However, in practice one has to contend with
(i)~confusion due to the overlap of low-redshift, low-order lines with
high-redshift, high-order lines, and (ii)~with an increased
uncertainty in the placement of the QSO's continuum. Both of these
difficulties are aggravated as one proceeds up the series towards
shorter wavelengths. For these reasons we will not consider any
\ion{H}{i} transitions that are of higher order than \lyb.

We now extend our simulated spectra by adding the region between a
QSO's \lyb\ and \lyg\ emission lines immediately bluewards of the
\lya\ forest. This region extends the redshift path for \lya\ lines by
a factor of $1.28$ towards lower redshifts and also contains \lyb\
lines in the redshift range $\lambda_\gamma / \lambda_\beta (1 +
\zqso) - 1 < z < \zqso$, where $\lambda_\gamma = 972.54$~\AA. Note,
however, that the real \lya\ line lists do not extend below the QSOs'
\lyb\ emission lines (cf.\ Table~\ref{rqsos}). Hence we are forced to
supplement the real \lyb\ absorption lines in this region with \lya\
lines drawn from the MC simulations.

\begin{figure}
\psfig{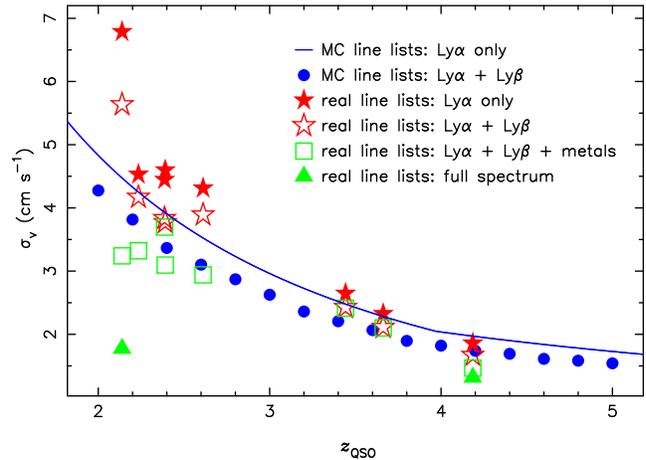}
\caption{Comparison of $\sv$ measurements derived from \lya-only line
  lists and various extended line lists covering additional spectral
  regions and including other absorption lines. The solid line and
  solid stars show, respectively, the results from the \lya-only MC
  and real line lists as in Fig.~\ref{svr_z}. The blue dots and open
  stars show the corresponding improved $\sv$ values that result from
  the addition of the \lyb\ forest. For the real line lists, the open
  squares show the effect of further adding the available metal lines
  in the \lya+\lyb\ forest region. Finally, for two of these QSOs, we
  show as solid triangles the outcome of using all lines accessible in
  existing spectra.}
\label{svbm_z}
\end{figure}

In Fig.~\ref{svbm_z} we compare the $\sv$ measurements derived from
the extended MC line lists (blue dots) to the \lya-only results (solid
line, same as in Figs.~\ref{sv_z} and \ref{svr_z}). We find that
including the \lyb\ forest improves $\sv$ by a factor of $0.88 \pm
0.006$. The corresponding comparison for the real line lists (open and
solid stars) yields improvement factors of $0.82$ to $0.92$, with a
mean of $0.88$, in very good agreement with the MC results.

\subsection{Metal lines}
\label{sv_m}
In addition to the \ion{H}{i} absorption, every high-redshift QSO
spectrum shows absorption lines from a number of other ions, such as
\ion{C}{iv}, \ion{Si}{iv} or \ion{Mg}{ii}. Although these metal lines
are far less numerous than the \ion{H}{i} lines, they are also much
narrower: their widths are of order a few\kms. In fact, many metal
lines are unresolved in current spectra and so the widths of the
narrowest lines are unknown. This suggests that metal lines may
considerably increase a spectrum's sensitivity to radial velocity
shifts and may hence supply valuable additional constraints on \zdot.

Could peculiar motions vitiate this supposition? This question is
difficult to answer with certainty because the exact origin of many of
the various classes of metal absorption lines is still under debate.
It is nevertheless evident that the structures responsible for the
absorption form an inhomogeneous set
\citep[e.g.][]{Churchill00,Churchill07} and represent a large range of
environments, depending on the absorbing ion, column density and
redshift. For example, low column density \ion{C}{iv} absorbers probe
the IGM \citep[e.g.][]{Simcoe04,Songaila06} while strong \ion{Mg}{ii}
lines are associated with galaxies (e.g.\ \citealp{Bergeron91};
\citealp*{Steidel94}; \citealp{Zibetti05}). However, we note that
absorbers associated with galaxies are generally found to have
absorption cross-sections on the order of tens of kpc
\citep[e.g.][]{Steidel95,Churchill00,Adelberger05}. For such distances
and for halo masses of $\sim$$10^{12} M_\odot$
\citep{Steidel94,Bouche04,Bouche06} one derives accelerations of a
$\mbox{few} \times 10^{-8}$~cm~s$^{-2}$, while absorber kinematics
point to peculiar velocities of no more than a few hundred\kms\
\citep[e.g.][]{Churchill96}. According to Appendix \ref{vpec},
peculiar motions of this magnitude are not problematic. The main
caveat here is the unknown fraction of {\em strong} metal absorbers
arising in starburst-driven outflows \citep[e.g.][see also Section
\ref{feedback}]{Adelberger03,Simcoe06,Bouche06} which must be expected
to experience much larger accelerations. However, it may be possible
to identify these systems kinematically \citep{Prochaska08} and hence
to exclude them from the analysis. In any case, in this section we
will proceed on the assumption that peculiar motions do not generally
invalidate the use of metal lines for a \zdot\ measurement.

Equation \eref{vp_distribution} gave a succinct parameterisation of
the properties of the \lya\ forest. Unfortunately, equivalents for all
the various metal line species do not exist, and hence we are unable
to gauge the effects of metal lines using MC simulations.

\citet{Kim01,Kim02} and \citet{Lu96}, from whom we obtained the real
\lya\ forest line lists used in the previous sections, also identified
metal lines in their spectra. Unfortunately, they only published
measured parameters for lines lying in the \lya\ forest region.
Although these lines will also have transitions elsewhere (which were
probably used in the identification process) the published line lists
do not provide a complete view of the metal line population outside of
this region.

To improve on this situation we went back to the spectra of two of our
QSOs, Q1101$-$264 and Q0000$-$26, and derived our own metal line lists
covering the entire spectral range available to us. In the case of
Q1101$-$264 the spectrum was the same as that used by \citet{Kim02},
whereas for Q0000$-$26 we used a UVES/VLT spectrum of similar quality
as the HIRES/Keck spectrum used by \citet{Lu96}. In the following we
will refer to these two line lists as the `complete' metal line lists.

Table~\ref{rmqsos} summarises the wavelength ranges that were searched
for metal lines and the total number of metal absorbers that were
found in each QSO spectrum. The simulated spectra are generated in
exactly the same way as in Section \ref{sv_lyb}, except that we now
add all available metal lines, and extend the spectra of Q1101$-$264
and Q0000$-$26 to the red limits given in Table~\ref{rmqsos}.

\begin{table}
\begin{center}
\begin{minipage}{6.6cm}
\caption{Observed lists of metal absorbers.}
\label{rmqsos}
\begin{tabular}{lcrc}
\hline
Name & $\lambda$ range$^a$ & $N_{\rm m}^b$ & Reference\\
\hline 
Q1101$-$264    & $3050$$-$$5765$, & $225$ & this work\\
               & $5834$$-$$8530$\\
J2233$-$606    & $3400$$-$$3850$ & $49$ & 2\\
HE1122$-$1648  & $3500$$-$$4091$ & $18$ & 1\\
HE2217$-$2818  & $3550$$-$$4050$ & $59$ & 2\\
HE1347$-$2457  & $3760$$-$$4335$ & $48$ & 1\\
Q0302$-$003    & $4808$$-$$5150$ & $5$ & 1\\
Q0055$-$269    & $4852$$-$$5598$ & $14$ & 1\\
Q0000$-$26     & $4300$$-$$6450$, & $100$ & this work\\
               & $7065$$-$$8590$,\\
               & $8820$$-$$9300$\\
\hline
\end{tabular}\\
References: 1 = \citet{Kim02}, 2 = \citet{Kim01}.\\
$^a$Wavelength range(s) searched for metal lines in \AA.\\
$^b$Number of metal absorbers, defined as the number of unique sets
of \{ion, $z$, $N$, $b$\}.
\end{minipage}
\end{center}
\end{table}

Fig.~\ref{svbm_z} shows the effect of the metal lines on $\sv$. The
green open squares show the result of only using the \lya+\lyb\ forest
region as in Section \ref{sv_lyb} but adding in all the available
metal lines in this region. Comparing this to the \ion{H}{i}-only
results of the previous section (open red stars) we find that the
metal lines improve $\sv$ by factors of $0.58$ to $0.99$, with a mean
of $0.85$. We point out that, strictly speaking, the derived $\sv$
values are only upper limits for the six line lists taken from the
literature because of their incomplete coverage of the \lyb\ forest
region. Indeed, the best improvement is achieved for one of the
complete line lists, Q1101$-$264 (the lowest redshift QSO), which has
a particularly rich metal absorption spectrum (cf.\
Table~\ref{rmqsos}).

Finally, for our two complete line lists we show as solid green
triangles the effect of also adding in the accessible spectral regions
redwards of the \lya\ forest. The additional metal lines further
improve the $\sv$ values of Q1101$-$264 and Q0000$-$26 by factors of
$0.55$ and $0.90$, respectively. With only two values it is obviously
impossible to draw a firm conclusion regarding the average improvement
offered by the metal lines redwards of the \lya\ forest. We therefore
choose to be conservative and adopt the larger of the two as a typical
value.

Summarising Sections \ref{sv_lyb} and \ref{sv_m}, the above
experiments have shown that the normalisation of equation \eref{sveq}
can be reduced by a factor of $0.88 \times 0.85 \times 0.9 = 0.67$ by
considering not just the \lya\ forest but all available absorption
lines, including metal lines, over the entire accessible optical
wavelength range down to a QSO's \lyg\ emission line. Hence we now
obtain:
\begin{equation}
\label{sveqm}
\sv = 1.35 \left(\frac{\mbox{S/N}}{2370}\right)^{-1} \!
\left(\frac{\nqso}{30}\right)^{-\frac{1}{2}} \!
\left(\frac{1 + \zqso}{5}\right)^{-1.7} \!\!\!\!\mbox{\cms}.
\end{equation}

\section{Multiple epochs}
\label{mult_epoch}
Fundamentally, a redshift drift experiment consists of simply
measuring the velocity shift between two spectra of the same QSO(s)
taken at two distinct epochs separated by some time interval $\Delta
t_0$. This is the view we took in the previous two sections where we
determined the fundamental photon-noise limit of the accuracy with
which this shift can be measured, and its scaling behaviour. However,
in practice this notion is too simplistic. First of all, it implicitly
assumes that the total integration time, $\ti$, required to achieve
the necessary S/N would be negligible compared to $\Delta t_0$, so
that the two epochs of observation are well-defined. As we will see in
the next section this assumption is not valid. Secondly, for a variety
of reasons it may be desirable to spread the observations more evenly
over the whole period $\Delta t_0$ instead of concentrating them in
just two more or less well-defined epochs at the interval's
endpoints. Hence the question arises how the accuracy of a redshift
drift experiment is affected by distributing the total available $\ti$
over multiple observing epochs within the interval $\Delta t_0$.

Let us assume then that observations take place at $N_e$ different
epochs, where the $j$th epoch is separated from the first by $\Delta
t_j$, so that $\Delta t_1 = 0$ and $\Delta t_{N_e} = \Delta t_0$. We
can straightforwardly generalise the framework developed in Section
\ref{sigv} by turning equation \eref{dvi} into a continuous equation
for the expected normalised flux in the $i$th pixel at time $\Delta
t$:
\begin{equation}
S_i(\Delta t) = S_{0i} + \frac{\d S_i}{\d \lambda} \; \lambda_i
\; \frac{\vdot_i}{c} \; \Delta t \equiv S_{0i} + m_i \Delta t.
\end{equation}
The idea is now to fit this linear relation to the observed fluxes
$S_{ji}$ at times $\Delta t_j$ with errors $\sigma_{ji}$, yielding an
estimate of the slope $m_i$ and hence of $\vdot_i$ for each
pixel. (Note that $S_{0i}$ is a nuisance parameter. It represents the
`true' flux at the first epoch as opposed to the observed value
$S_{1i}$.) The maximum likelihood estimator for $m_i$ is
\begin{equation}
m_i = \frac{\overline{S_i \Delta t} - \overline{S_i} \;
  \overline{\Delta t}}{\overline{\Delta t^2} - \overline{\Delta t}^2},
\end{equation}
where the bar denotes the weighted average over all epochs:
\begin{equation}
\overline{x} = \frac{\sum_{j=1}^{N_e} x \;
\sigma_{ji}^{-2}}{\sum_{j=1}^{N_e} \sigma_{ji}^{-2}}.
\end{equation}
The variance of $m_i$ is given by
\begin{equation}
\label{sigmi}
\sigma_{m_i}^2 = \left[\sum_j \sigma_{ji}^{-2} \left(\overline{\Delta
t^2} - \overline{\Delta t}^2\right)\right]^{-1}.
\end{equation}
With $m_i$ and its variance in place we can write down the equivalent
of equation \eref{sigvi}:
\begin{equation}
\label{sigvdoti}
\sigma^2_{\vdot_i} = \left (\frac{c}{\lambda_i \frac{\d S_i}{\d
      \lambda}}\right)^2 \left[ \sigma_{m_i}^2 \; + \;
      \frac{m_i^2}{\left(\frac{\d S_i}{\d \lambda}\right)^2} \;
      \sigma_{S^\prime_i}^2\right],
\end{equation}
which in turn allows us to compute $\vdot$ averaged over all pixels
and its error, $\sigma_{\vdot}$, corresponding to equations
\eref{deltav} and \eref{sve}. Finally, we re-define $\sv \equiv
\sigma_{\vdot} \, \Delta t_0$. This new version of $\sv$ now includes
the effect of multiple observing epochs and an arbitrary distribution
of the total integration time among them. It is straightforward to
show that for $N_e = 2$ one recovers exactly the original $\sv$ of
Section \ref{sigv}.

We are now in a position to amend equations \eref{sveq} and
\eref{sveqm} to include the effect of multiple epochs. Recall that
these scaling relations were derived for the case of $N_e = 2$ and for
equal S/N in the spectra of both epochs. Since the variance of the
normalised flux in pixel $i$ scales as the inverse of the integration
time we now write $\sigma_{ji}$ as
\begin{equation}
\sigma_{ji}^2 = \sigma_i^2 \; \frac{0.5}{f_j},
\end{equation}
where $f_j$ is the fraction of the total $\ti$ used at the $j$th epoch
($\sum_j f_j = 1$), and $\sigma_i$ denotes the flux error (in the
$i$th pixel) that one would obtain if half of the total $\ti$ were
used. Further defining $\Delta t_j \equiv h_j \Delta t_0$, we can
re-write equation \eref{sigmi} as:
\begin{eqnarray}
\lefteqn{\sigma_{m_i}^2 = \frac{2 \sigma_i^2}{\Delta t_0^2} \; \left\lbrace 4
  \left[\sum_j h_j^2 f_j - \left(\sum_j h_j f_j \right)^2\right]
  \right\rbrace^{-1}} \nonumber\\
& = & \!\!\!\sigma_{m_i}^2(N_e = 2, f_1 = f_2 = 0.5) \; 
g^2(N_e, h_{1 \ldots N_e}, f_{1 \ldots N_e}).
\end{eqnarray}
The first term above is just the variance of $m_i$ that one obtains in
the case of $N_e = 2$ and equal splitting of $\ti$. The second term is
a `form factor' that only depends on the distribution of $\ti$ within
$\Delta t_0$. Again, it is straightforward to show that $g(N_e = 2,
f_1 = f_2 = 0.5) = 1$. Since the form factor is the same for all
pixels, and since $\sigma_{m_i}^2$ is the dominant term in equation
\eref{sigvdoti}, the sought-after modification of the $\sv$ scaling
relation amounts to simply applying the form factor:
\begin{eqnarray}
\label{sveqmne}
\lefteqn{\sv = 1.35 \left(\frac{\mbox{S/N}}{3350}\right)^{-1} 
\left(\frac{\nqso}{30}\right)^{-\frac{1}{2}}
\left(\frac{1 + \zqso}{5}\right)^{-1.7}} \nonumber\\
& & \mbox{} \times g(N_e, f_{1 \ldots N_e}) \mbox{\cms}.
\end{eqnarray}
Note that the symbol `S/N' now refers to the {\em total} S/N per
object accumulated over all epochs, in contrast to equations
\eref{sveq} and \eref{sveqm} where it referred to the S/N achieved in
each of two epochs.

Note also that we have dropped the dependence of the form factor on
$h_{1 \ldots N_e}$ by considering every night within the period
$\Delta t_0$ as a potential epoch. This fixes $N_e$ and $h_j = (j-1)
/ (N_e - 1)$, while $f_j$ is constrained to lie in the range $0 \le
f_j \le l/\ti$, where $l$ is the length of a night (which we will
assume to be $9$~h on average). Thus we find:
\begin{equation}
g(N_e, f_{1 \ldots N_e}) = \frac{N_e - 1}{2} \left[\sum_{j=1}^{N_e-1} j^2
f_j - \left(\sum_{j=1}^{N_e-1} j f_j\right)^2\right]^{-\frac{1}{2}}\!\!.
\end{equation}

Clearly, $f_j$ will be $0$ for most nights. Nevertheless, there are
obviously a large number of different possible distributions for the
$f_j$s. The best distributions are those that are symmetric and peaked
towards the endpoints of $\Delta t_0$. A flat distribution, with equal
observations taking place on $n$ equally spaced nights, results in $g
= \sqrt{3 \, (n-1)/(n+1)} \approx 1.7$ for $n \gg 1$. Thus the
otherwise quite desirable arrangement of observing at a more or less
constant rate throughout the period $\Delta t_0$ comes with a rather
severe penalty attached. A priori, it is difficult to estimate the
best $g$ value that can be realistically achieved in practice. From
now on we will assume, perhaps somewhat arbitrarily, that all
observations occur as much as possible towards the beginning and end
of $\Delta t_0$ with the constraint that the observing rate averaged
over some intermediate timescale of, say, a month cannot exceed $1/3$,
i.e.\ that no more than a third of any month's telescope time is used
for the redshift drift experiment. Depending on the ratio of $\ti$ and
$\Delta t_0$ this results in $g$ values of $\sim$$1.1$. Essentially,
this configuration simply shortens the effective length of the
experiment by the amount of time it takes to complete the observations
at either end of $\Delta t_0$.

\section{Can we collect enough photons?}
\label{photons}
In Sections \ref{sims}, \ref{ext_spec} and \ref{mult_epoch} we learnt
what S/N ratio is required to achieve a given sensitivity to radial
velocity shifts using QSO absorption spectra. In a photon-noise
limited experiment the attainable S/N depends only on four quantities:
the brightness of the source, the size of the telescope's collecting
area, the total integration time and the total efficiency. By `total
efficiency' we mean the ratio of the number of detected
photo-electrons to the number of source photons at the top of the
atmosphere, i.e.\ it comprises atmospheric absorption and all losses
occurring in the combined telescope/instrument system, including
entrance aperture losses and the detector's quantum efficiency.

In this Section we will investigate in detail the 5-dimensional
parameter space that is spanned by the above four quantities and
redshift, in order to determine whether a feasible combination exists
that would allow a meaningful \zdot\ measurement.

\subsection{S/N formula}
We begin by writing down the relation between the S/N per pixel and
the above four parameters for the photon-noise limited case:
\begin{equation}
\label{sn}
\frac{\mbox{S}}{\mbox{N}} = 700 \left[ \frac{Z_X}{Z_r} \; 10^{0.4 (16 -
    m_X)} \; \left( \frac{D}{42 \, \mbox{m}} \right)^2 \; \frac{\ti}{10
    \, \mbox{h}} \; \frac{\epsilon}{0.25} \right]^\frac{1}{2},
\end{equation}
where $D$, $\ti$ and $\epsilon$ are the telescope diameter, total
integration time and total efficiency, $Z_X$ and $m_X$ are the zeropoint
and apparent magnitude of the source in the $X$-band, respectively,
and $Z_r = (8.88 \times 10^{10})$~s$^{-1}$~m$^{-2}$~$\mu$m$^{-1}$ is
the AB zeropoint \citep{Oke74} for an effective wavelength of
$6170$~\AA\ [corresponding to the Sloan Digital Sky Survey (SDSS)
$r$-band]. The normalisation of the above equation assumes a pixel
size of $0.0125$~\AA\ (see Section \ref{genspec}) and a central
obscuration of the telescope's primary collecting area of $10$ per
cent. $D = 42$~m corresponds to the Baseline Reference Design for the
European ELT \citep[E-ELT;][]{Gilmozzi07}.

\subsection{High-redshift QSOs}
The photon flux from QSOs is of course not a free parameter that can
be varied at will. Instead we will have to content ourselves with what
will be offered by the population of real QSOs known at the time of a
hypothetical \zdot\ experiment. Here we do not wish to speculate on
possible future discoveries of QSOs and hence we will restrict
ourselves to the ones known already today. In the following we will
extract a list of potential targets for a \zdot\ experiment from
existing QSO catalogues. For each candidate target QSO we will need a
reliable magnitude that can be used to estimate its photon flux, as
well as its redshift.

The largest QSO catalogue with reliable, homogeneous photometry and
redshifts currently available is the fourth edition of the SDSS Quasar
Catalog \citep{Schneider07}. Being based on the fifth data release of
the SDSS, it yields $16\,913$ QSOs with $\zqso \ge 2$. The catalogue
provides PSF magnitudes in the $ugriz$ bands which we do {\em not}
correct for Galactic extinction (as is appropriate for S/N
calculations). Since we are interested in the continuum flux we will
use, for each QSO, the magnitude of the bluest filter that still lies
entirely redwards of the QSO's \lya\ emission line. Specifically, for
objects with $\zqso < 2.2$ we will use the $g$-band magnitude; for
$2.2 \le \zqso < 3.47$ the $r$-band; for $3.47 \le \zqso < 4.61$ the
$i$-band; and for $4.61 \le \zqso$ the $z$-band. We then apply a small
correction to the selected magnitude to transform the observed flux to
that expected at the centre of the \lya\ forest assuming a power-law
spectral shape of the form $f_\nu \propto \nu^{-0.5}$
\citep{Francis93}.

Unfortunately, the SDSS catalogue does not cover the whole sky. The
largest collection of QSOs covering the {\em entire} sky is the 12th
edition of the Catalogue of Quasars and Active Nuclei recently
published by \citet{VeronCetty06}, which contains many additional
bright QSOs not included in the SDSS catalogue. However, since the
V{\'e}ron catalogue is a compilation of data from many different
sources its photometry is very inhomogeneous and cannot readily be
converted to a photon flux. We will dispense with this inconvenience
by matching the V{\'e}ron catalogue to the all-sky SuperCOSMOS Sky
Survey \citep[SSS;][]{Hambly01c}. Although the photographic SSS
photometry is not endowed with reliable absolute calibrations either,
at least it is homogeneous and covers three bands ($B_{\rm J}$, $R$
and $I$), allowing us to synthesise approximate SDSS magnitudes.

We proceed by first selecting all QSOs from the V{\'e}ron catalogue
with $\zqso \ge 2$ and not flagged as unreliable, resulting in
$21\,895$ objects. For each of these we then identify the nearest SSS
object, allowing a maximum separation of $5$~arcsec, and retrieve the
corresponding SSS catalogue data. $938$ V{\'e}ron objects have no SSS
counterpart, presumably because their coordinates are unreliable. We
then use those $11\,744$ objects that are in common to the SDSS and
combined V{\'e}ron-SSS catalogues to derive linear relations between
the SDSS and SSS magnitudes, allowing for zeropoint offsets and colour
terms. Such relations are reasonable representations of the data and
we find that the distributions of the residuals have r.m.s.\ values of
less than $0.3$~mag in all cases. Finally, we purge the common objects
from the V{\'e}ron-SSS catalogue and use the above relations to
synthesise an SDSS magnitude in the appropriate band (see above) for
each of the remaining QSOs.

\begin{figure}
\psfig{file=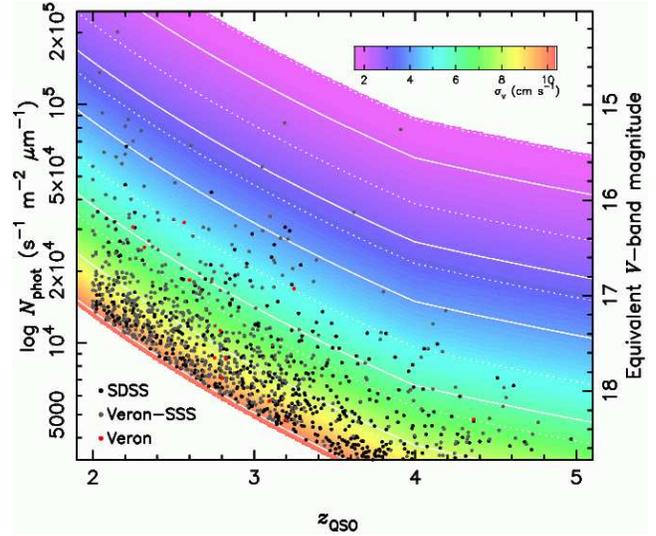,width=\columnwidth}
\caption{The dots show the known, bright, high-redshift QSO population
  (separated by sub-sets as indicated, see text) as a function of
  redshift and estimated photon flux at the centre of the \lya\
  forest. Along the right-hand vertical axis we have converted the
  photon flux to a corresponding Johnson $V$-band magnitude. The
  background colour image and solid contours show the value of $\sv$
  that can be achieved for a given photon flux and redshift, assuming
  $D = 42$~m, $\epsilon = 0.25$ and $\ti = 2000$~h. The contour levels
  are at $\sv = 2, 3, 4, 6, 8$ and $10$\cms. The dotted contours show
  the same as the solid ones, but for $D = 35$~m or, equivalently, for
  $\epsilon = 0.17$ or $\ti = 1389$~h.}
\label{m_z}
\end{figure}

For those QSOs in the initial V{\'e}ron catalogue that have no match
in the SSS, or which are missing an SSS band needed to synthesise the
required SDSS magnitude, we will simply use the $V$ or $R$-band
magnitude as listed in the V{\'e}ron catalogue, provided it is
non-photographic.

In summary, the final combined sample of $25\,974$ QSOs is constructed
from three sub-sets: (i)~SDSS; (ii)~objects with redshifts from the
V{\'e}ron catalogue and photometry from the SSS (converted to the SDSS
system); and (iii)~objects where both the redshifts and the photometry
are taken from the V{\'e}ron catalogue. We remind the reader that the
quality and reliability of the photometry decreases rapidly from (i)
to (iii).

\subsection{Achievable radial velocity accuracy}
In Fig.~\ref{m_z} we plot our QSO sample, split by the above sub-sets,
in the $\np$-$\zqso$ plane, where $\np$ is a QSO's photon flux at the
top of the atmosphere and at the centre of the QSO's \lya\ forest, as
implied by the appropriate magnitude described above. Using equations
\eref{sveqmne} and \eref{sn} and assuming values for $D$, $\epsilon$
and $\ti$ we can calculate, for any given combination of $\np$ and
$\zqso$, the value of $\sv$ that would be achieved if {\em all} of the
time $\ti$ were invested into observing a single QSO with the given
values of $\np$ and $\zqso$. The background colour image and solid
contours in Fig.~\ref{m_z} show the result of this calculation, where
we have assumed $D = 42$~m, $\epsilon = 0.25$ and $\ti = 2000$~h. Note
that we have included both the improvement of $\sv$ afforded by the
\lyb\ forest and the metal lines as well as the deterioration caused
by spreading $\ti$ over a $0.9$~yr period at either end of a $\Delta
t_0 = 20$~yr interval.

From Fig.~\ref{m_z} we can see that, although challenging, a
reasonable measurement of $\dot z(z)$ is within reach of a $42$-m
telescope. There exist a number of QSOs that are bright enough and/or
lie at a high enough redshift to provide reasonable values of
$\sv$. We find $18$ objects at $\sv < 4$\cms\ and $5$ objects at $\sv
< 3$\cms, with good coverage of the redshift range $2$--$4$. One
object even gives $\sv = 1.8$\cms. However, for a smaller telescope
with $D = 35$~m the number of objects with $\sv < 4$\cms\ reduces to
only $7$ (cf.\ dotted contours).

\begin{figure}
\psfig{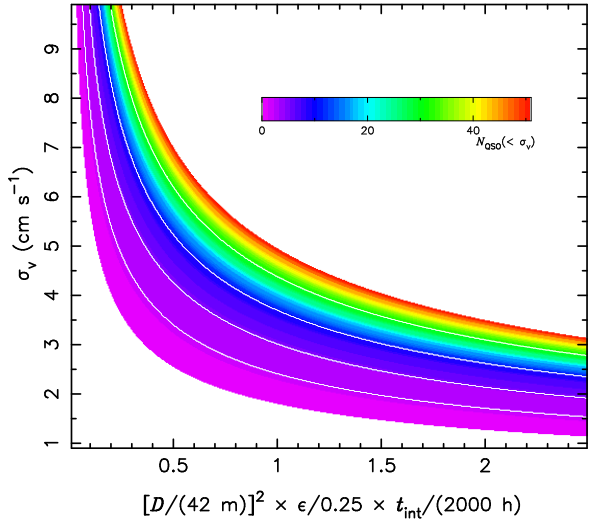}
\caption{The colour image and the contours show the number of QSOs for
  which the $\sv$ value on the ordinate or better can be achieved for a
  given combination of telescope size, efficiency and integration time.
  The contour levels are at $\nqso = 3,5,10$ and $30$.}
\label{sigv_O}
\end{figure}

In Fig.~\ref{sigv_O} we show more comprehensively how the number of
QSOs with $\sv$ smaller than a given value depends on the telescope
parameters and integration time, which we summarise into a single
`normalised observational setup parameter', $O$:
\begin{equation}
O = \left( \frac{D}{42 \, \mbox{m}} \right)^2 \;
    \frac{\epsilon}{0.25} \; \frac{\ti}{2000 \, \mbox{h}}
\end{equation}
(cf.\ equation \ref{sn}). For a given value of $O$ the colour image
and contours in Fig.~\ref{sigv_O} show the number of QSOs that would
give a $\sv$ equal to or smaller than the value along the ordinate if
all of $\ti$ was spent on any one of them. For example, if we wanted
to be able to choose our targets from $30$ QSOs bright enough and/or
at a high enough redshift to be able to achieve $\sv = 3$\cms\ or
better on each object individually, then we would require $O \approx
2.1$. Note however, that the ordinate of Fig.~\ref{sigv_O} does {\em
not} give the overall value of $\sv$ for a \zdot\ experiment using the
best $\nqso$ targets and setup $O$. The reason is of course that the
total value of $\sv$ of such an experiment depends on how the total
integration time is split up among the $\nqso$ targets (see below).

\subsection{Target selection and observing strategies}
The question of how to select targets for a \zdot\ experiment depends
on what exactly one wishes to achieve. For example, we could simply
rank the candidate targets by their achievable $\sv$ for a fixed
observational setup $O$, as suggested by Fig.~\ref{m_z}, and select
the objects with the smallest values. This would be the correct
selection strategy if the goal were to obtain the smallest possible
overall $\sv$ of the combined sample. Another possible goal might be
to maximise the significance with which $\dot z \neq 0$ can be
detected. In this case we would select the targets with the largest
values of $|\vdot| / \sigma_{\vdot}$, which would give much larger
weight to the high redshift objects and would practically deselect all
objects at $\zqso \la 2.5$ where the redshift drift is too small to be
detected. Yet another option would be to maximise the sensitivity to
$\ol$ (at a fixed, given value of $\om$) by selecting the objects with
the largest values of $\frac{\d \vdot}{\d \ol} / \sigma_{\vdot}$. It
is clear that the choice of selection method will also depend on the
possible adoption of priors from other cosmological observations.

Once the exact goal and hence the target selection strategy has been
defined we must also choose the number of targets to include in the
experiment. Including more than just the `best' target (according to a
given selection strategy) is desirable for several practical reasons,
including the ability to observe for as large a fraction of the total
available telescope time as possible (to ease scheduling), and to be
able to identify any potential directional systematics caused by the
atmosphere, the telescope, the instrument, the calibration procedures
or the transformation of redshifts to the cosmological reference
frame. Obviously though, the more objects are included into the
experiment the worse the final result will be (in the absence of
systematics) because observing time will have to be redistributed from
the `best' object to the less suited ones.

Finally, we must also decide how to divide the total available
integration time among the objects included in the experiment. Again,
there are several possibilities, including: the time could be split
equally among the targets or in such a way as to ensure equal S/N,
equal $\sv$ or an equal value of the selection parameter. (Note that
in practice it is likely that operational and scheduling constraints
would limit the available choices.)

From the above it is clear that there are a large number of different
possibilities of implementing a \zdot\ experiment. In particular, the
decision on what precisely the objective of such an experiment should
be does not appear to be straightforward. We will now focus on three
alternatives which may be perceived to be representative of three
different approaches.

1. The simplest possible goal is to aim for the most precise \zdot\
   measurement possible, i.e.\ the smallest overall $\sv$. This may be
   considered a `pure experimentalist' approach where virtually no
   prior observational information or theoretical expectation is used
   to (mis-)guide the design of the experiment.

2. Another approach is to emphasise the most basic and yet most unique
   (and perhaps most captivating) feature of the redshift drift
   experiment, which is being able to {\em watch}, literally and in
   real time, the Universe changing its rate of expansion.  In this
   case the aim is to prove the existence of a dynamical redshift
   drift effect and hence to measure a non-zero value of \zdot\ with
   the highest possible significance.\footnote{The value of this
   approach is perhaps best appreciated by picturing the PI of a
   future \zdot\ experiment angrily muttering ``And yet it moves!''
   under her/his breath as s/he leaves a seminar room after having
   tried to convince a sceptical audience that (i)~an effect has
   indeed been observed and (ii)~that it is not an instrumental
   artifact.} However, selecting targets according to this approach
   requires the adoption of, and hence dependence on, a specific model
   for $\dot z(z)$, and is therefore not free of prior assumptions, in
   contrast to our first example.

3. As explained in the introduction, the discovery and the unknown
   physical source of the acceleration of the universal expansion
   provide a strong motivation for any observation seeking to
   determine $H(z)$, and this is also true for the direct and
   dynamical method of measuring \zdot. From this point of view the
   goal should be to place the strongest possible constraints on the
   acceleration. Translating this goal to a target selection strategy
   again requires a model of $\dot z(z)$, including the acceleration.

\begin{figure}
\psfig{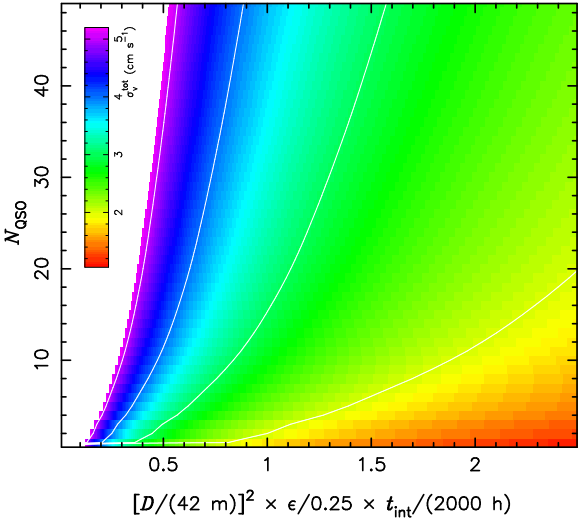}
\caption{The colour image and the contours show the final, overall
  value of $\sv$ achieved by targeting the $\nqso$ objects with the
  best individual $\sv$ values (cf.\ Fig.~\ref{m_z}) and by employing
  a given combination of telescope size, efficiency and total
  integration time. The total integration time is split equally among
  the targets. The contour levels are at $\sv^{\rm tot} = 2,3,4$ and
  $5$\cms.}
\label{nqso_O}
\end{figure}

\subsection{Simulations of example measurements and constraints on
  model parameters}
We now proceed to implement each of the above approaches on the QSO
sample shown in Fig.~\ref{m_z} in order to illustrate what can be
achieved with each of them. However, we stress that the above
approaches and the specific implementations below are only example
strategies, and that many variants, refinements and completely
different procedures are possible.

We begin with the simplest case of selecting targets by the value of
their achievable $\sv$, which can be read off Fig.~\ref{m_z} for each
object. Having chosen the best $\nqso$ targets by this criterion we
can compute the total, overall value of $\sv$ that can be achieved for
this sample and for a given observational setup, $O$:
\begin{equation}
\sv^{\rm tot} = \left[\sum_i^{\nqso} \sv^{-2}(z_{{\rm QSO}, i},
m_i, f_{{\rm obj},i} \, O)\right]^{-\frac{1}{2}},
\end{equation}
where $f_{{\rm obj},i}$ is the fraction of the total integration time
allocated to the $i$th object. In Fig.~\ref{nqso_O} we plot $\sv^{\rm
tot}$ as a function of $\nqso$ and $O$, where we have chosen to
distribute the time equally among all objects (i.e.\ $f_{{\rm obj},i}
= \nqso^{-1}$). Distributing the time to give equal S/N instead
(which may be operationally desirable) leads to slightly better values
of $\sv^{\rm tot}$ when only a few QSOs are targeted. The reason is of
course that the object with the best $\sv$ is only the seventh
brightest target (cf.\ Fig.~\ref{m_z}) and so it is allocated
relatively more time. However, for $\nqso \ga 15$ this advantage is
lost and equal time allocation produces $\sv^{\rm tot}$ values that
are smaller by $\sim$$10$ per cent compared to equal S/N allocation.

From Fig.~\ref{nqso_O} we can see that an overall value of $\sv^{\rm
tot} \approx 2$--$3$\cms\ is well within reach of an ELT, even when
tens of objects are targeted for the experiment. The steepness of the
contours also indicate that $\sv^{\rm tot}$ does not depend very
sensitively on $\nqso$, at least for $\nqso \ga 10$.

\begin{figure}
\psfig{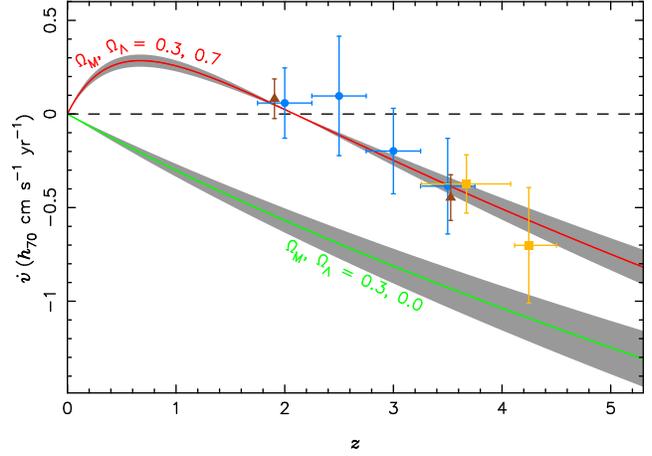}
\caption{The three sets of `data' points show MC simulations of the
  redshift drift experiment using the three different example
  implementations discussed in the text. In each case we have assumed
  an observational setup of $O = 2$ and we plot as `data' points the
  `observed' values and error bars of the velocity drift $\vdot$,
  expected for a total experiment duration of $\Delta t_0 = 20$~yr and
  for standard cosmological parameters ($h_{70} = 1$, $\om = 0.3$,
  $\ol = 0.7$). For a given QSO we use the centre of the \lya\ forest
  as the redshift of the $\vdot$ measurement. Blue dots: selection by
  $\sv$, $\nqso = 20$ (binned into four redshift bins), equal time
  allocation. Yellow squares: selection by $|\vdot| / \sigma_{\vdot}$,
  $\nqso = 10$ (in two redshift bins), equal time allocation. Brown
  triangles: selection by best combined constraint on $\ol$, $\nqso =
  2$, optimal time distribution. The solid lines show the expected
  redshift drift for different parameters as indicated, and $h_{70} =
  1$. The grey shaded areas result from varying $H_0$ by $\pm
  8$\kms~Mpc$^{-1}$.}
\label{dv_z}
\end{figure}

For further illustration we now perform a MC simulation of a redshift
drift experiment using this target selection strategy. We
(arbitrarily) choose $\nqso = 20$ and $O = 2$ so that the overall
accuracy of the experiment is $\sv^{\rm tot} = 2.34$\cms\ (cf.\
Fig.~\ref{nqso_O}). Furthermore we assume that the observations span a
time interval of $\Delta t_0 = 20$~yr and that the true $\dot z(z)$ is
given by our standard cosmological model. The result is presented in
Fig.~\ref{dv_z} where we show as blue dots one realisation of the
`observed' velocity drifts and their errors along with the input model
(red solid line). Since the selected $20$ targets cover the redshift
range $2.04 \le \zqso \le 3.91$ quite homogeneously we have binned the
measurements into four equally sized redshift bins.

By construction these points represent the most precise measurement of
\zdot\ that is possible with a set of $20$ QSOs (using equal time
allocation) and $O = 2$. However, since many of the selected QSOs lie
near the redshift where $\dot z = 0$ (for the assumed model) the
redshift drift is only detected with an overall significance of
$\mathcal{S} = \overline{|\vdot|} / \sigma_{\vdot}^{\rm tot} =
1.4$, where $\overline{|\vdot|}$ is the weighted mean of the
absolute values of the expected velocity drifts.

This can be improved upon by choosing our second approach and
selecting targets by the largest value of $|\vdot| /
\sigma_{\vdot}$. This quantity is a strongly increasing function of
redshift and this selection strategy results in quite a different set
of objects: the best $20$ targets according to this criterion include
only $3$ of the objects previously selected. For $O = 2$, $\Delta t_0
= 20$~yr and equal time allocation as before, the combined overall
significance of this sample is $\mathcal{S} = 2.7$ (assuming our
standard model of \zdot\ is correct). The yellow squares in
Fig.~\ref{dv_z} show the result of a MC simulation using this setup
and implementation, except that we use $\nqso = 10$ (in two redshift
bins) which gives a slightly better significance of $\mathcal{S} =
3.1$. To reach $\mathcal{S} \ge 4$ we need to further reduce $\nqso$
to $3$, or instead increase $O$ to $3.4$ or $\Delta t_0$ to $25.4$~yr.

Finally, we turn to our third approach and the question of how to best
select targets to constrain the acceleration of the expansion and what
can be achieved in this respect with our sample of known QSOs. As in
the previous case the answer will depend on what to expect for the
expansion history and in particular for the acceleration. For the
purpose of the following discussion we will again assume our standard
cosmological model.

The simplest thing we can do to constrain the acceleration is to
unambiguously detect its existence, i.e.\ to measure $\dot z > 0$ with
the highest possible significance. This implies (i)~that we need a
$\vdot$ measurement at $z < z_0$, where $z_0$ is defined by
$\vdot(z_0) = 0$, and (ii)~that target selection should proceed by the
largest value of $\vdot / \sv$, which indeed favours the lowest
available redshifts. However, even if we use only the single best
object by this criterion and assume a generous $\Delta t_0 = 25$~yr
then a $2\sv$ detection of $\vdot > 0$ would still require an
unfeasible $O = 9.5$. The reason for not being able to do better is of
course our inability to access the \lya\ forest at $z \approx 0.7$
where $\vdot$ is the largest.

Let us analyse the situation more systematically by switching to the
parameter space of our cosmological model (which is 3-dimensional
since we will {\em not} assume spatial flatness). In this parameter
space our goal of detecting the acceleration translates to proving
that the deceleration parameter $q_0 = \frac{\om}{2} - \ol$ is $<
0$. However, in our model the acceleration is due to a cosmological
constant, and so for simplicity we will instead pursue the slightly
easier goal of proving the existence of a cosmological constant, i.e.\
of placing a positive lower limit on $\ol$ after having marginalised
over $\om$ and $H_0$.

Consider the constraint of a single measurement of $\vdot = \Delta v /
\Delta t_0$ at some $z$ (i.e.\ provided by some QSO), which happens to
be precisely equal to its expected value. Obviously, a single data
point cannot constrain all three parameters at the same time but only
some 2-dimensional surface. Solving equations \eref{zdot} and
\eref{fried} for $\ol$ we find that this surface is defined by:
\begin{equation}
\label{ol}
\ol = \frac{(1+z)^2}{(1+z)^2 - 1} \; \left[z \, \om + 1 - \left(1 -
    \frac{\vdot}{c H_0}\right)^2 \right].
\end{equation}
Since $\vdot$ is assumed to be equal to its expectation value this
constraint surface is entirely defined by $z$ and it must include the
point $(\om, \ol, h_{70}) = (0.3, 0.7, 1)$. Furthermore, the
measurement error $\sigma_{\vdot}$ endows the surface with some small
thickness. Hence we can see that each QSO in our sample [representing
a ($z,\sigma_{\vdot}$) pair] defines its own constraint surface. The
question is whether there exists some combination of some sub-set of
these surfaces such that the projection of the combined constraint
onto the $\ol$-axis does not include the region $\ol \le 0$. If so,
what is the largest significance level at which $\ol = 0$ can be
excluded?

\begin{figure}
\psfig{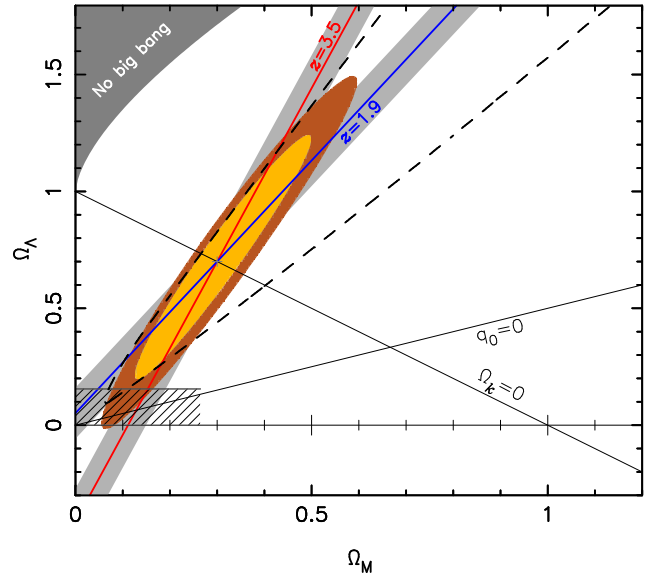}
\caption{Expected constraints in the $\ol$-$\om$ plane from two
  measurements of the redshift drift at two different redshifts as
  indicated, assuming $O = 2$ and $\Delta t_0 = 20$~yr. The two
  targets and the split of the observing time among them were chosen
  to give the best lower limit on $\ol$. The red and blue solid lines
  and the grey shaded bands show the individual constraints provided
  by each of the two objects assuming a fixed $h_{70} = 1$. The
  coloured ellipses show the joint $68$ and $90$ per cent confidence
  regions that result from combining the two measurements,
  marginalising over $H_0$ using an external prior of $H_0 = (70 \pm
  8)$\kms~Mpc$^{-1}$. The hashed region indicates the $95$ per cent
  lower limit on $\ol$. The dashed contour shows the $68$ per cent
  confidence region that results from using a flat prior on $H_0$ and
  just marginalising over the range $0 \le H_0 \le
  140$\kms~Mpc$^{-1}$. Flat cosmologies and the boundary between
  current de- and acceleration are marked by solid black lines. The
  dark shaded region in the upper left corner designates the regime of
  `bouncing universe' cosmologies which have no big bang in the past.}
\label{ol_om}
\end{figure}

To address this problem let us fix $h_{70} = 1$ for now, so that a
given constraint surface turns into a band in the $\ol$-$\om$ plane
(cf.\ the grey shaded bands in Fig.~\ref{ol_om}). Consider first the
case $\nqso = 1$. For a fixed $H_0$, equation \eref{ol} simply gives a
linear relation between $\ol$ and $\om$ which has a non-zero slope for
all $z > 0$. It follows that the projection of the constraint band
onto the $\ol$-axis will be infinite unless we can place some kind of
limit on $\om$. Since the slope is positive we require a lower limit
on $\om$ in order to obtain a lower limit on $\ol$. Rather
uncontroversially we can impose $\om > 0$. Applying this limit to
equation \eref{ol} we find that a single $\vdot$ measurement can place
a positive lower limit on $\ol$ if the intercept of the $\ol$-$\om$
relation, $b$, is significantly larger than $0$. The error on $b$ is
given by the width of the constraint band along the $\ol$-direction:
\begin{equation}
\label{dol}
\Delta b = \Delta \ol = \frac{\d \ol}{\d \vdot} \; \sigma_{\vdot} =
\frac{2}{c H_0^2} \; \frac{(1+z) H(z)}{(1+z)^2 - 1} \;
\frac{\sv}{\Delta t_0}.
\end{equation}
and the object giving the best lower limit on $\ol$ is the one with
the largest value of $b / \Delta b$. However, we find that even the
best object by this criterion does not deliver a significant lower
limit on $\ol$. Hence we find that not only are we unable to detect
$\vdot > 0$ with any significance but it is also impossible to prove
$\ol > 0$ with a {\em single} $\vdot$ measurement.

However, this conclusion rests entirely on the weak lower limit on
$\om$ that we used ($\om > 0$). The larger this limit the stronger
also the lower limit on $\ol$. In other words, a given $\vdot$
measurement translates to a different constraint on $\ol$ depending on
what we assume for $\om$. Hence, even if we cannot detect $\vdot > 0$
with high significance it is nevertheless possible to place a
significant lower limit on $\ol$. All we need is a stronger constraint
on $\om$.

The point is that this constraint can be supplied by the \zdot\
experiment itself, without having to resort to an external prior, by
means of a second measurement at {\em high} redshift (as opposed to
at $z \la z_0$). The idea is to combine the measurements from two
objects (or two groups of objects) that are selected in different
ways, such that their constraint bands in the $\ol$-$\om$ plane make
as large an angle as possible while also being as narrow as possible
(cf.\ Fig.~\ref{ol_om}). The first group is selected to have shallow
constraint lines, implying low redshifts ($z \la z_0$) and supplying a
lower (but by itself insufficient) limit on $\ol$ as discussed above,
while the second group is selected to have steep constraint lines,
implying high redshifts (cf.\ equation \ref{ol}) and supplying a lower
limit on $\om$. This is illustrated in Fig.~\ref{ol_om} where the red
and blue lines and the grey shaded areas show the constraint bands
from two $\vdot$ measurements at low and high redshifts as
indicated. If the angle between them is large enough and if the
individual $\sv$s are small enough then the combination of the two
constraints (in a $\chi^2$ sense) results in an ellipse in the
$\ol$-$\om$ plane whose lower edge excludes $\ol = 0$.

Hence we find that we need at least $\nqso = 2$. However, recall that
so far we have kept $H_0$ fixed. Let us now reconsider the above as a
function of $H_0$. First of all, we note that the constraint line
resulting from the $\vdot$ measurement at $z \approx z_0$ is very
insensitive to the value of $H_0$ (because $\vdot \approx
0$). Secondly, for the high redshift measurement the constraint line
shifts upwards in Fig.~\ref{ol_om} as $H_0$ increases. Hence the
centre of the joint ellipse (where the constraint lines cross) moves
from $(\om, \ol) = (0.3, 0.7)$ approximately along the low-$z$
constraint line towards the lower left of Fig.~\ref{ol_om}. However,
at the same time the extent of the ellipse decreases because a larger
$H_0$ predicts a larger redshift drift and hence results in stronger
constraints on $\om$ and $\ol$ for fixed measurement errors. The net
result is that the lower edge of the joint ellipse and hence the
obtainable lower limit on $\ol$ are relatively insensitive to
$H_0$. As we will see more quantitatively below, a third $\vdot$
measurement to constrain $H_0$ is therefore not necessary.

Thus we find that the strategy of selecting two groups of targets as
described above should provide the best possible lower limit on $\ol$,
even after marginalising over $H_0$. However, when trying to express
this approach in equations in order to proceed with the target
selection one realises that the selection of the two groups cannot in
fact be separated, making an analytical procedure impractical. Hence we
resort to the simplest case of only one object per group and simply
try all possible combinations of two objects in our sample. For any
given combination we also determine empirically the optimal split of
the total integration time among the two objects. In this way we find
that the best lower limit on $\ol$ is achieved for the two very bright
objects in Fig.~\ref{m_z} at $\zqso = 2.15$ and $3.91$, with the
latter object receiving $0.3$ of the total integration time. Mock
$\vdot$ measurements for these objects are shown as brown triangles in
Fig.~\ref{dv_z}.

In Fig.~\ref{ol_om} we present the expected constraints from these two
objects in the $\ol$-$\om$ plane, assuming $O = 2$ and $\Delta t_0 =
20$~yr. The red and blue solid lines show the individual constraint
lines for a fixed $h_{70} = 1$ (equation \ref{ol}) and the grey shaded
areas around them are the error bands that correspond to the
respective values of $\sv$ for each object (equation \ref{dol}). Note
that the optimal time split (in the sense of giving the best lower
limit on $\ol$) is such that the two bands are of roughly equal width.

The coloured ellipses in Fig.~\ref{ol_om} show the joint $68$ and $90$
per cent confidence regions that result from the combination of the
two $\vdot$ measurements. Here we have assumed that an external
constraint places $1$$\sigma$ limits of $\pm8$\kms~Mpc$^{-1}$ on $H_0$
\citep{Freedman01} before marginalising over this parameter. By
comparing the error bars on the brown triangles in Fig.~\ref{dv_z}
with the grey shaded region around the red line we can see that this
uncertainty in $H_0$ is small relative to the errors on $\vdot$. Hence
the ellipses in Fig.~\ref{ol_om} are almost those that would be
obtained for a fixed $H_0$.

Projecting the joint constraints onto the vertical axis we find that
the $95$ per cent lower limit on the cosmological constant is $\ol >
0.16$ (indicated by the hashed region in Fig.~\ref{ol_om}), while $\ol
= 0$ is excluded at the $98.2$ per cent confidence level. Note that we
can also exclude $q_0 = 0$ at a similar level of $97.6$ per
cent. These numbers do not depend sensitively on the adopted prior on
$H_0$, for reasons explained above. Doubling the error on $H_0$ to
$\pm 16$\kms~Mpc$^{-1}$ has almost no effect at all. Even if we use a
flat prior and simply marginalise over the range $0 \le H_0 \le
140$\kms~Mpc$^{-1}$ we can still reject $\ol = 0$ at $98.1$ per cent
confidence, although the $95$ per cent lower limit now reduces to
$0.08$. For a shorter experiment duration of $\Delta t_0 = 15$~yr (and
the flat $H_0$ prior) we can still exclude $\ol = 0$ at $93.5$ per
cent confidence.

\begin{figure}
\psfig{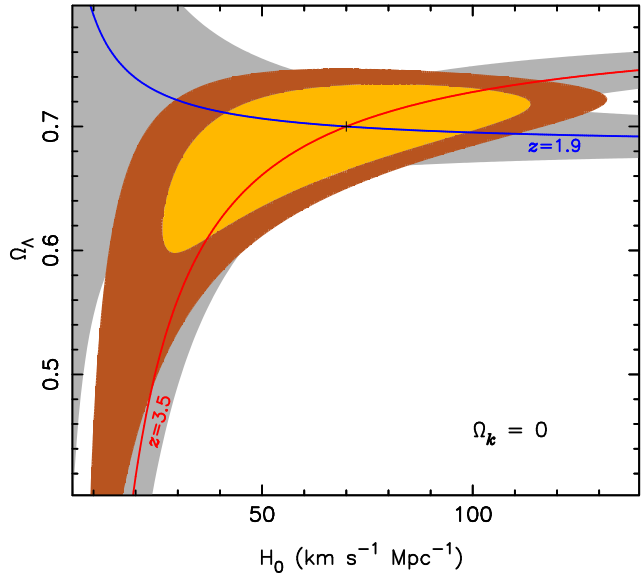}
\caption{Expected constraints in the $\ol$-$H_0$ plane for flat
  cosmologies assuming $O = 2$ and $\Delta t_0 = 20$~yr and using the
  same two redshift drift measurements as in Fig.~\ref{ol_om}. The red
  and blue solid lines and the grey shaded bands show the individual
  constraints provided by each of the two objects. The coloured
  ellipses show the joint $68$ and $90$ per cent confidence regions
  that result from combining the two measurements.}
\label{ol_H0}
\end{figure}

Thus we conclude that, assuming general relativity to be valid, an
ELT will enable us to confirm the existence of a cosmological constant
using purely dynamical evidence -- by measuring the redshift drift in
QSO absorption lines over a period of $\sim$$20$~yr -- {\em without
assuming flatness, using any other external cosmological constraints
or making any other astrophysical assumptions whatsoever}.

Combining the constraints of a \zdot\ experiment with those from other
cosmological observations would of course lead to significantly
stronger results. A detailed examination of this issue is, however,
beyond the scope of this paper. We will let it suffice to make two
general points: (i)~the \zdot\ constraints will be unique and
therefore new in the sense that they cannot be provided by any other
cosmological observation. \zdot\ probes $H(z)$ in a redshift range
that is inaccessible to other methods (such as SNIa, weak lensing,
BAO) and this will remain true even in the ELT era. In terms of
characterising $H(z)$, \zdot\ measurements at $z \ga 2$ will nicely
complement the data at $z \la 2$ provided by SNIa surveys. In the
$\ol$-$\om$ plane these datasets offer similar but nevertheless
distinct constraints. (ii)~The datasets that are most complementary to
the redshift drift in the $\ol$-$\om$ plane are those that constrain
the geometry of the Universe, such as the fluctuation power spectrum
of the CMB, which can break the remaining degeneracy in the \zdot\
constraints. In fact, the WMAP 3-year data constrain the quantity $\ol
+ 0.72 \, \om$ \citep{Spergel07}, so that the CMB degeneracy line is
almost exactly orthogonal to the \zdot\ constraints shown in
Fig.~\ref{ol_om}. Combining these constraints would lead to individual
$2$$\sigma$ errors in $\ol$ and $\om$ of $\sim$$0.08$ and
$\sim$$0.06$, respectively.

Finally, let us consider models in which the Universe is {\em exactly}
flat ($\ok = 0$). The parameter space of these models is 2-dimensional
and a single $\vdot$ measurement results in a constraint line given by
\begin{equation}
\ol = \frac{(1+z)^2}{(1+z)^3 - 1} \; \left[1+z - \left(1 -
    \frac{\vdot}{c H_0}\right)^2 \right].
\end{equation}
In Fig.~\ref{ol_H0} we show the constraints in the $\ol$-$H_0$ plane
offered by the two $\vdot$ measurements considered above ($O = 2$,
$\Delta t_0 = 20$~yr). Clearly, $H_0$ is not constrained in any useful
way and both substantially higher and lower redshift measurements of
similar or better quality would be required to do so. On the other
hand, $\ol$ is constrained quite well. Marginalising over $H_0$ we
find a $2$$\sigma$ range of $[0.42,0.74]$, and $\ol = 0$ is excluded
at $98.7$ per cent confidence. For a shorter experiment duration of
$\Delta t_0 = 15$~yr we can still exclude $\ol = 0$ at $94.7$ per cent
confidence. Note that these latter values are not very much higher
than in the general case above. The difference is of course that in
the flat case the best fit along the $\ol = 0$ axis is obtained for an
unfeasibly low $H_0$ value of $h_{70} = 0.1$, while in the general
case this value is $1$. Hence, for flat cosmologies any external $H_0$
prior has a strong impact on the lower limit one can place on $\ol$,
in contrast to the general case. For example, applying the relatively
weak prior that $H_0$ is known to within $\pm 16$\kms~Mpc$^{-1}$
results in the rejection of $\ol = 0$ at $>99.99$ per cent confidence
for any $\Delta t_0 > 9$~yr or, alternatively, for $\Delta t_0 =
15$~yr and any $O > 0.5$.

\subsection{Constraints on alternative cosmological models}
So far we have only considered the standard cosmological model, i.e.\
general relativity with a stress-energy tensor dominated by cold
(dark) matter and a cosmological constant (\lcdm). Needless to say,
industrious theorists have proposed numerous alternative explanations
for the observed acceleration of the universal expansion. Broadly
speaking, these models either suggest alternative forms of dark energy
or modify the theory of gravity. In Appendix \ref{altcos} we give a
brief overview of two examples for each of these two classes of
models. These are quintessence, generalised Chaplygin gas, Cardassian
expansion and the DGP brane-world model. Despite fundamental
differences, some generic features are shared by all four. Obviously,
they all `predict' late-time acceleration. They also agree that the
early evolution of the Universe is indistinguishable from
matter-dominated expansion in general relativity (so as not to spoil
the successes of the standard model in explaining early structure
formation). Furthermore, all of them have at least as many free
parameters as \lcdm. In fact, three of the four models considered here
include \lcdm\ phenomenologically as a special case, i.e.\ as a
sub-space of the models' parameter space. With these properties in
mind, and given that \lcdm\ can successfully reproduce SNIa, CMB and
BAO data, it is perhaps not too surprising that the alternative models
can also be adjusted to fit these datasets, resulting in best-fit
parameter values consistent with \lcdm\ \citep[e.g.][]{Davis07}.

\begin{figure}
\psfig{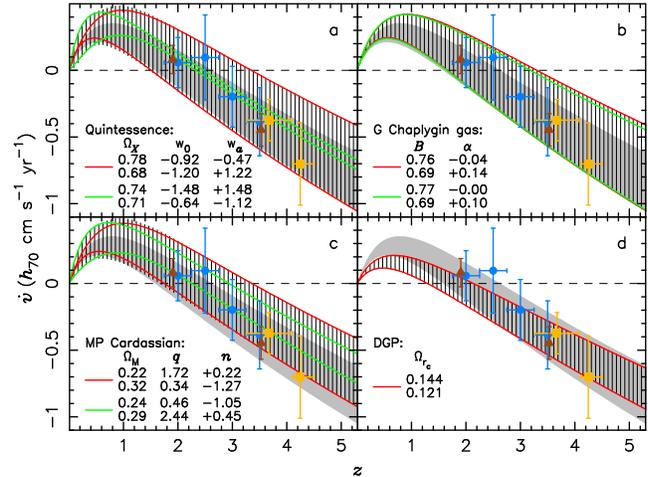}
\caption{Comparison of the predictions for the redshift drift by
  non-standard cosmological models. For simplicity, we impose flatness
  and fix $h_{70} = 1$ in all cases. (a)~At a given redshift the
  vertically hashed region shows the range of \zdot\ values predicted
  by quintessence, allowing the model's three free parameters to vary
  within the joint $95.4$ per cent confidence region imposed by SNIa,
  CMB and BAO data \citep{Davis07}. The green and red solid lines show
  the predictions for those parameter combinations that produce the
  most extreme allowed \zdot\ values at $z = 0.3$ and $4$,
  respectively.  $(w_0, w_a) = (-1, 0)$ corresponds to flat
  \lcdm. (b--d)~Same as (a) for the other models as indicated. For the
  Chaplygin gas and Cardassian models $\alpha = 0$ and $(q, n) = (1,
  0)$ correspond to flat \lcdm, respectively. In each panel we also
  show, for comparison, the predictions of the flat \lcdm\ model with
  $0.65 \le \ol \le 0.74$ as the grey shaded region. We also reproduce
  the three sets of simulated measurements from Fig.~\ref{dv_z}.}
\label{dv_z_ns}
\end{figure}

Given this state of affairs we do not think it justified to study
these alternative models, in this paper, to the same level of detail
as we did for \lcdm\ in the previous section. The name of the game
would be to ask, for each model, how much of the non-\lcdm-like
parameter space could be excluded by a redshift drift experiment. This
would best be done by considering the \zdot\ constraints jointly with
those from other relevant datasets. Instead, we will do something much
simpler here. As already mentioned above, \citet{Davis07} have fit all
four of our alternative models to current SNIa, CMB and BAO data. The
idea is now to ask, for each model, which values of \zdot\ are still
`allowed' given the constraints already placed on the model's
parameters by the \citeauthor{Davis07} analysis. Comparing these
allowed ranges with the simulated \zdot\ measurements of the previous
section will give a good impression of what can be achieved.

In Fig.~\ref{dv_z_ns} we plot this allowed \zdot\ range as the
vertically hashed region for all four models considered here along
with the predictions for some specific parameter combinations as
indicated. Here we have assumed $\Delta t_0 = 20$~yr. In order to
reduce the number of free parameters we are only considering flat
models with $H_0$ fixed at $h_{70} = 1$. For each model the plotted
\zdot\ range corresponds to the joint $95.4$ per cent confidence
region in the space spanned by the model's remaining free
parameters. The grey shaded region in each panel shows the flat \lcdm\
model (which has only one free parameter since we have fixed $\ok$ and
$H_0$), varying $\ol$ in the range $[0.65,0.74]$. This corresponds to
the $95.4$ confidence range obtained from the two redshift drift
measurements discussed in the previous section. Note that the hashed
and grey shaded regions are not `centred' on each other. The reason
is that we chose to centre the grey shaded region on our fiducial
value of $\ol = 0.7$, whereas the ranges of the equivalent parameters
in panels (a)--(c) are centred on $0.73$. For reference, we also
reproduce, in each panel, the simulated data points from
Fig.~\ref{dv_z}.

In panels (a)--(c) the hashed regions are somewhat larger than the
grey shaded regions and the error bars on the `data' points,
indicating that a redshift drift experiment will be able to
significantly reduce the sizes of the allowed regions in these models'
parameter spaces. The specific parameter combinations that we chose
for the solid lines are those that produce the most extreme allowed
\zdot\ values at $z = 0.3$ (green lines) and at $z = 4$ (red
lines). For the quintessence and Cardassian models we can see that
these two sets of lines have very different parameter values. Indeed,
the two parameter combinations that produce the extreme \zdot\ values
at low redshift predict very similar values at high redshift, and vice
versa (although to a lesser extent). This again highlights the
complementarity between $H(z)$ constraints at low and high
redshifts. Notice, however, that the situation is different in panel
(b). In the case of generalised Chaplygin gas one set of parameters
produces the maximum (or minimum) allowed \zdot\ values at more or
less all redshifts. The reason is that the region in this model's
2-dimensional parameter space allowed by the \citeauthor{Davis07}
analysis is in fact nearly 1-dimensional. This is of course exactly
true for the DGP model (panel d) which has only one free parameter to
begin with. Notice, however, that in this case $\vdot(z)$ has a
somewhat different slope. In contrast to the other models, the DGP
prediction for \zdot\ cannot be made arbitrarily close to the \lcdm\
prediction and from Fig.~\ref{dv_z_ns} we can see that the differences
are of roughly the same size as the expected \zdot\ measurement
errors.

We defer a detailed investigation of these issues to a future
analysis. Here we simply conclude that the increased redshift range of
$H(z)$ measurements afforded by a redshift drift experiment will
significantly add to our capability of discriminating between
alternative cosmological models.

\section{Conclusions}
\label{conclusions}
In this paper we have appraised the prospects of using an ELT to
detect and characterise the cosmological redshift drift, \zdot, by
means of a $\sim$two decade-long spectroscopic monitoring campaign of
high redshift QSO absorption lines. This experiment would be unique
among cosmological observations in that it directly probes the global
dynamics of the Robertson-Walker metric, thereby offering a novel and
truly independent path to the expansion history of the Universe. We
summarise our results as follows:

1. We confirm the earlier proposition \citep{Loeb98} that the \lya\
   forest is the most promising target for a redshift drift
   experiment. We have explicitly examined the peculiar motions of the
   gas responsible for the \lya\ forest in a hydrodynamical simulation
   and found their effects negligible. We have also assessed the
   effects of variations of the physical properties of the absorbing
   gas over the timescale of the experiment and found them similarly
   insignificant. 

2. Under the assumption of photon-noise limited observations the
   accuracy, $\sv$, to which a putative radial velocity shift between
   two spectra of the same object can be determined depends chiefly on
   the number and sharpness of the relevant spectral features and the
   spectra's S/N. The number density of the absorption lines that
   constitute the \lya\ forest depends strongly on redshift. Using
   extensive Monte Carlo simulations we have hence derived a
   quantitative relation between the $\sv$ of a pair of \lya\ forest
   spectra on the one hand, and the spectra's S/N and the background
   QSO's redshift on the other hand (equation \ref{sveq}).

3. Apart from the \lya\ forest, all QSO spectra also display higher
   order \ion{H}{i} lines and a variety of metal absorption
   lines. Using the \ion{H}{i} \lyb\ lines in addition to the \lya\
   lines, as well as all available metal lines within and to the red
   of the \lya\ forest significantly improves a given spectrum's
   sensitivity to radial velocity shifts by a factor of $0.67$
   (equation \ref{sveqm}).

4. In practice it will not be possible to measure the redshift drift
   by simply comparing just two observations of some set of objects
   obtained at two well-defined epochs separated by some $\Delta
   t_0$. In reality, the total observing time required for the
   experiment will be spread over many nights (= epochs) whose
   distribution within $\Delta t_0$ (now defined as the interval
   between the first and last observations) will be subject to many
   practical constraints. In Section \ref{mult_epoch} we have
   calculated the relevant `form factor' by which $\sv$ increases (or,
   equivalently, by which the effective $\Delta t_0$ decreases) as a
   result of a given observing time distribution. For realistic
   distributions we find factors of $\sim$$1.1$ to $1.7$.

6. The detailed implementation of a redshift drift experiment will
   depend on the exact objectives one hopes to achieve. We have
   elaborated three different possible top-level goals and we have
   worked out the specific target selection strategies required to
   tailor a \zdot\ experiment to each of these. Applying these
   selection schemes to a comprehensive list of high redshift QSOs
   {\em already known today} we use our final $\sv$ scaling relation
   to predict an ELT's capability to measure \zdot. For example,
   selecting those $20$ targets that deliver the greatest measurement
   accuracy we find that a $42$-m ELT could achieve a total, overall
   accuracy of $2.34$\cms\ in $4000$~h of observing time
   (Fig.~\ref{nqso_O}). If we additionally assume that our standard
   cosmological model correctly predicts the size of the redshift
   drift then we will be able to prove its existence (i.e.\ measure a
   non-zero value) at a significance of $3.1\sigma$ in $\Delta t_0 =
   20$~yr (cf.\ Fig.~\ref{dv_z}).

   Unfortunately however, due to the fact that the \lya\ forest is
   only accessible from the ground for $z \ga 1.7$ it will {\em not}
   be possible to directly measure $\dot z > 0$ (the hallmark of past
   accelerated expansion) from any {\em one} of the QSOs in our sample
   with any significance, unless the transition from decelerated to
   accelerated expansion occurred significantly earlier than predicted
   by our standard model. However, even if the standard model's
   prediction for \zdot\ is correct an ELT will nevertheless be able
   to provide unequivocal proof of the existence of past acceleration
   by obtaining an upper limit on \zdot\ at high redshift and a lower
   limit at $z \approx 2$ where the transition from $\dot z < 0$ to
   $>0$ is expected to occur (cf.\ Fig.~\ref{dv_z}). For the above
   experiment parameters these limits will be strong enough to prove
   that \zdot\ must become positive below $z \approx 2$. We stress
   that the validity of this proof only requires that gravity can be
   described by a metric theory and that the Universe is homogeneous
   and isotropic on large scales. It is entirely independent of any
   specific theory of gravity (such as general relativity) and does
   not require any assumptions regarding the spatial geometry of the
   Universe. It is also independent of other cosmological observations
   and does not rely on any other astrophysical assumptions
   whatsoever. Hence, a \zdot\ experiment carried out with an ELT will
   arguably provide the most direct evidence of acceleration possible.

   In the context of general relativity dominated by cold dark matter
   and a cosmological constant [$(\om, \ol) = (0.3, 0.7)$] a \zdot\
   experiment over $\Delta t_0 = 20$~yr using $4000$~h of observing
   time on a $42$-m ELT will be able to exclude $\ol = 0$ with $98.1$
   per cent confidence, providing a $2$$\sigma$ lower limit of $0.08$
   (cf.\ Fig.~\ref{ol_om}). Additionally assuming spatial flatness
   results in a $2$$\sigma$ range of $0.42 < \ol < 0.74$
   (Fig.~\ref{ol_H0}).

7. Finally, we point out that a \zdot\ experiment constrains $H(z)$
   over a redshift range that is inaccessible to other
   methods. Redshift drift measurements at $z \ga 2$ would therefore
   complement current and future SNIa and BAO surveys. Adding \zdot\
   to the suite of modern cosmological tests would thus provide
   additional leverage in constraining the parameters of both \lcdm\
   and alternative cosmological models (cf.\ Fig.~\ref{dv_z_ns}).

We stress that all the results summarised above were derived under the
assumption that the uncertainty on radial velocity shift measurements
will be dominated by photon noise. The next step is to understand to
what extent this assumption can be realised in practice. To this end
we need to (i)~draw up a list of all potential sources of error, both
astrophysical and instrumental in nature, (ii)~quantitatively assess
their impact on a \zdot\ measurement, and (iii)~where necessary and
possible identify appropriate hardware solutions, calibration and/or
analysis techniques that will reduce their impact below the level of
the photon noise. Although we have already addressed some of the most
obvious astrophysical sources of noise in Section \ref{lyaf}, a
realistic and comprehensive assessment of the total error budget is a
non-trivial task -- in particular because of point~(iii) above --
which we defer to a future paper.

One of the potential error sources that will require careful attention
is the fact that QSO continua are to some extent variable. The
flux-difference method we described in Section \ref{sigv} to extract
the redshift drift signal from the data is clearly very sensitive to
any errors in determining the change of the continuum. Similarly, it
is also quite susceptible to any residual uncertainties in the sky
subtraction, flat-fielding and stray light correction. Clearly, a
flux-difference type of approach would not be the method of choice for
a real-life \zdot\ experiment -- a conclusion we had already reached
in Section \ref{sigv} for an entirely different reason. However, other
methods of extracting the redshift drift signal are likely to be less
sensitive to these issues. As already suggested in Section
\ref{ext_spec} we believe that the most promising approach is to
simultaneously model the absorption in all spectra of a given object
with Voigt profiles, including the \zdot\ of each component (or a
model thereof) as an additional free parameter. The drawback of this
method is that a complex absorption feature can usually be modelled in
more than one way, none of which necessarily has to correspond to
reality, and this could represent an additional source of error for a
\zdot\ measurement. In any case, the question of how to deal with QSO
continuum variability remains to be investigated.

Another issue we have not yet studied is the effect of the evolution
of structure along the line of sight to a distant QSO. Consider two
photons whose redshift difference we wish to measure. Along their way
to us both will encounter various gravitational potential wells. Any
evolution of these potentials between the passage of the first and
second photon will cause them to follow slightly different paths,
perturbing their redshift difference. This effect will have to be
studied with high time resolution simulations of cosmic structure
formation.

The quest for photon-noise limited radial velocity shift measurements
also places rather exacting requirements on the spectrograph to be
used for a redshift drift experiment, in particular with respect to
its stability and wavelength calibration. However, an instrument
concept study investigating these and other issues has already been
carried out \citep[CODEX;][]{Pasquini05}, concluding that the
technical challenges are demanding but not insurmountable. One of the
novel concepts identified by this study is the use of a `laser
frequency comb' for wavelength calibration. This system provides a
series of uniformly spaced, narrow lines whose {\em absolute}
positions are known a priori with relative precision better than
$10^{-12}$ (see \citealp{Murphy07} for details). Note that an absolute
wavelength calibration of this precision would allow the combination
of data from different ELTs and would guarantee the legacy value of
the data by enabling future astronomers to measure \zdot\ over much
longer timescales.

In deriving our results we have also used the magnitudes and redshifts
of the high-redshift QSO population as it is known today. Note that we
have used {\em all} known QSOs, irrespective of their declination. In
exchange, so to speak, we have refrained from speculating on future
discoveries of bright, high-redshift QSOs. It is clear that existing
QSO catalogues do not yet cover the entire sky uniformly with the same
completeness, even at the bright magnitudes we are interested in
here. Indeed, only one third of the objects shown in Fig.~\ref{m_z}
lie in the south. Furthermore, since colour selection is one of the
most popular methods of QSO candidate identification it is likely that
our current sample is incomplete in the redshift range $2.4 \la z \la
3$ where the colours of QSOs are very similar to those of main
sequence stars \citep[e.g.][]{Richards02}. Hence it is not
unreasonable to believe that at least a few bright QSOs still remain
to be discovered, especially in the south. Considering that a \zdot\
experiment would benefit greatly from {\em any} new discoveries we
propose that a complete, systematic survey for bright, high-redshift
QSOs be carried out once the site for an ELT has been chosen.

In addition, we propose that all potential target QSOs should be
monitored photometrically. Not only would this remove the
uncertainties of our present photometry (note that we had to rely on
photographic photometry from the SuperCOSMOS Sky Survey for most of
the brightest objects in Fig.~\ref{m_z}) but it would also allow us to
account for QSO variability in the target selection. If the monitoring
frequency were high enough (as can be expected from possible future
facilities like the Large Synoptic Survey Telescope) then these data
could also be used to time the spectroscopic observations of a given
target to coincide with a peak in the object's lightcurve.

We close by observing that any QSO spectra collected as part of a
\zdot\ experiment could also be used to address numerous other
important issues, including the variability of the fine-structure
constant and the primordial abundance of deuterium. Given the
substantial observing time requirements for a \zdot\ experiment it is
clear that these synergies will play an important role in any
realistic attempt to establish a future ELT redshift drift campaign.

\section*{Acknowledgements} 
Numerical computations were performed on the COSMOS supercomputer at
the Department of Applied Mathematics and Theoretical Physics in
Cambridge. COSMOS is a UK-CCC facility which is supported by HEFCE and
PPARC.

\bibliographystyle{mn2e}


\appendix

\section{Redshift drift including peculiar motion}
\label{vpec}
Equation \eref{zdot} for the redshift drift is only correct if both
the observer and the source are comoving. In reality, both will have
peculiar velocities and accelerations due to local gravitational
potential wells. The purpose of this Appendix is to provide an exact
calculation of the effects of peculiar motion on the observed redshift
drift.

Our own peculiar motion can be broken down into several components,
such as the Earth's motion with respect to the solar system's
barycentre, the solar system's rotation within the Galaxy, and the
Galaxy's movement with respect to the Hubble flow. All of these
components are or will soon be known with high precision so that the
observations can be corrected accordingly \citep{Pasquini06}. Hence we
will not consider the observer's peculiar motion any further.

For a given object at large distance it is in general not possible to
determine its peculiar motion or to correct its observed redshift for
it. However, the peculiar motions of widely separated objects are
expected to be uncorrelated. Hence, when using several such objects to
measure $\dot z$, their peculiar motions are only expected to
introduce a random noise component, but will not cause any systematic
bias. To assess the relevance of peculiar motions for any given
implementation of a \zdot\ experiment one must compare the noise
induced by the peculiar motions of the experiment's targets with other
sources of error (ideally only photon noise).

\begin{figure}
\psfig{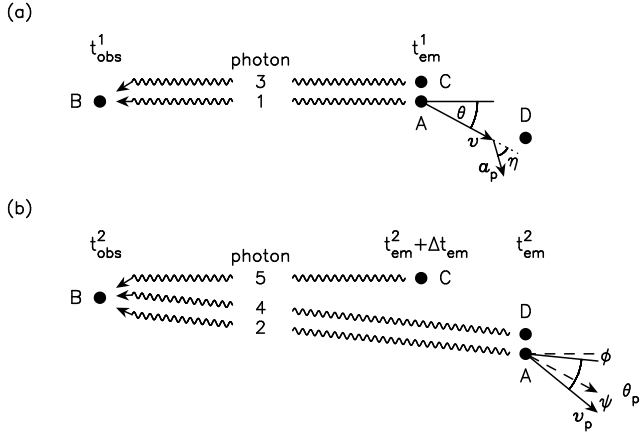}
\caption{Cartoon illustrating various photons, their emitters and
  observers, as well as their times of emission and observation (see
  text for details). We also show various angles that define the
  peculiar motion of object A: the angle $\theta$ in panel (a) is the
  angle between the peculiar velocity $\bmath{v}$ and the LOS, while
  $\eta$ is the angle between $\bmath{v}$ and the peculiar
  acceleration projected into the plane of the drawing,
  $\bmath{\ap}$. The acceleration $\bmath{a}$ itself makes an angle
  $\zeta$ with this plane (not shown). In panel (b) $\bmath{v}$ has
  acquired a component perpendicular to the plane of the drawing
  (making an angle $\xi$), and so we only show its projected
  component, $\bmath{\vp}$, as well as its angle $\tp$ with the
  LOS. Compared to panel (a) $\bmath{v}$ has shifted by an angle
  $\psi$ in the plane of the drawing while the direction of the LOS
  has changed by the angle $\phi$.}
\label{cartoon}
\end{figure}

We now derive the expression for the observed redshift drift in the
presence of peculiar velocity and acceleration. Consider a photon
emitted at time $\tem$ by a source moving with velocity $\bmath{v}$ with
respect to its local Hubble flow (object A in Fig.~\ref{cartoon}), and
received by a comoving observer at time $\tobs$ (object B in
Fig.~\ref{cartoon}). Clearly, the observed redshift of this photon
travelling from A to B is given by
\begin{equation}
\label{zab}
1+\zobs(\tobs, \tem) = [1 + \zd(\tem)] \, [1 +
z(\tobs, \tem)].
\end{equation}
Here, $z$ is the cosmological redshift as before, and $\zd$ is the
(special relativistic) Doppler shift of the emitter:
\begin{equation}
1 + \zd(\tem) = \frac{1 + \cos [\theta(\tem)] \frac{v(\tem)}{c}}{\sqrt{1 -
[\frac{v(\tem)}{c}]^2}} = \gamma \, (1 + \cos \theta \vc),
\end{equation}
where $\theta$ is the angle between the emitter's peculiar velocity
and the outward line of sight (LOS) from the observer (see
Fig.~\ref{cartoon}). The redshift drift is observed as the redshift
difference between the photons labelled as 1 and 2 in
Fig.~\ref{cartoon}, and has two terms:
\begin{eqnarray}
\label{dzobsdt}
\lefteqn{\frac{\zobs(\tobs^2, \tem^2) - \zobs(\tobs^1,
\tem^1)}{\tobs^2 - \tobs^1} \; \approx \; \frac{\d \zobs}{\d \tobs}}
\nonumber\\ 
& = & \frac{\d \zd}{\d \tobs} \; (1 + z) + (1 + \zd) \; 
\frac{\d z}{\d \tobs}.
\end{eqnarray}
We can therefore identify the following, separable effects of peculiar
motions on the process of putting data points on Fig.~\ref{dzdt}: (i)
The Doppler shift in equation \eref{zab} introduces an uncertainty on
where to place a given object along the $x$-axis of Fig.~\ref{dzdt}.
Uncertainties in the placement along the $y$-axis are caused by (ii)
peculiar acceleration, (iii) the Doppler shift in the second term of
equation \eref{dzobsdt} and (iv) the change in the comoving
coordinates of the source. This last point can in turn be separated
into two distinct effects: the decrease of $\theta$ (due to parallax)
and the change of the distance between the observer and the
source. The latter implies that the $\d z/\d\tobs$ appearing in
equation \eref{dzobsdt} is {\em not} the same as that of equation
\eref{zdot}, which is only valid for a constant $\chi$.

This is illustrated by photons 3 and 4 in Fig.~\ref{cartoon}. Photon 3
is emitted at $\tem^1$ by the {\em comoving} object C that is at the
same location as the moving object A at time $\tem^1$, while photon 4
is emitted by comoving object D just as A is passing by at time
$\tem^2$. Photons 3 and 4 are received by B at the same times as
photons 1 and 2, respectively (i.e.\ $\tobs^1$ and
$\tobs^2$). Clearly, the redshift difference between these two photons
is the cosmological shift we are seeking:
\begin{equation}
\d z = z_{\rm D \rightarrow B}(\tobs^2, \tem^2) - z_{\rm C \rightarrow
B}(\tobs^1, \tem^1)
\end{equation}
To calculate $\d z$ we make use of photon 5 in
Fig.~\ref{cartoon}. Like photon 3 this photon is sent from C to B, although at
a different time. Since C and B are comoving objects, the redshift
difference of photons 3 and 5 is given by equation \eref{zdot}. We
have:
\begin{eqnarray}
\label{dzcosmo}
\d z & = & z_{\rm D \rightarrow B}(\tobs^2, \tem^2) - 
z_{\rm C \rightarrow B}(\tobs^2, \tem^2 + \Delta \tem) \nonumber\\
& & \mbox{} + z_{\rm C \rightarrow B}(\tobs^2, \tem^2 + \Delta \tem) -
z_{\rm C \rightarrow B}(\tobs^1, \tem^1) \nonumber\\
& \approx & -\frac{\partial z}{\partial \tem} \; \Delta \tem + 
\frac{\d z_{\vert\chi}}{\d \tobs} \; \d \tobs \nonumber\\
& = & (1 + z) \; H(\tem) \; \Delta \tem \nonumber\\
& & \mbox{} + [(1+z) \; H(\tobs) - H(\tem)] \; \d \tobs \nonumber\\
& = & (1+z) \; H(\tobs) \; \d \tobs \nonumber\\
& & \mbox{} - (1+z) \; H(\tem) \; (\d t_{\rm C} - \Delta \tem) \nonumber\\
& = & (1+z) \; H(\tobs) \; (\tobs^2 - \tobs^1) \nonumber\\
& & \mbox{} - (1+z) \; H(\tem) \; (\tem^2 - \tem^1),
\end{eqnarray}
where $\Delta \tem$ is the time required by photon 4 to cover the
comoving radial coordinate difference, $\Delta \chi$, between C and D
as seen from B.  $\d t_{\rm C}$ is the time difference between the
emission of photons 3 and 5 by C, so that $\d \tobs = (1+z) \, \d
t_{\rm C}$. The last equality of equation \eref{dzcosmo} holds by
construction. Using
\begin{equation}
c \Delta \tem = a(\tem) \Delta \chi = \cos \theta v \, (\tem^2 - \tem^1)
\end{equation}
one obtains
\begin{equation}
\frac{\tem^2 - \tem^1}{\tobs^2 - \tobs^1} = [(1+z) \; (1 + \cos \theta
\vc)]^{-1}
\end{equation}
and hence
\begin{eqnarray}
\frac{\d z}{\d \tobs} & = & (1 + z) \, H(\tobs) - (1 + \cos \theta
\vc)^{-1} \, H(\tem) \nonumber\\
& = & (1 + z) \, H(\tobs) - \frac{\gamma}{1 + \zd} \, H(\tem).
\end{eqnarray}

Turning now to the first term of equation \eref{dzobsdt} the change in
the Doppler shift is given by:
\begin{equation}
\d \zd = \frac{\d \zd}{\d \tem} \; (\tem^2 - \tem^1)
\end{equation}
so that
\begin{equation}
\frac{\d \zd}{\d \tobs} = \frac{\d \zd}{\d \tem} \; [(1+z) \; (1 + \cos \theta
\frac{v}{c})]^{-1},
\end{equation}
where $\d \zd/\d \tem$ is the rate of change of the Doppler shift as
measured by a local comoving observer along the LOS to B. Using a
prime to denote differentiation with respect to $\tem$, we have
\begin{equation}
\label{dzddt}
\zd\p = \frac{v\p}{c} \; (1+\zd) \left(\frac{\cos\theta}{1 + \cos
  \theta \vc} + \gamma^2 \vc \right) - \gamma \sin\theta \; \vc \;
  \theta\p.
\end{equation}

In general, the peculiar acceleration, $\bmath{a}$, can have any
orientation with respect to $\bmath{v}$. We denote the direction of
$\bmath{a}$ by the angles $\zeta$ and $\eta$. The former is the angle
between $\bmath{a}$ and the plane spanned by $\bmath{v}$ and the
observer (i.e.\ the plane of Fig.~\ref{cartoon}). The latter is the
angle between $\bmath{\ap}$, the projection of $\bmath{a}$ into this
plane, and $\bmath{v}$ [see Fig.~\ref{cartoon}(a)]. Hence we have
\begin{equation}
v\p = \cos\eta \; \ap = \cos\eta \, \cos\zeta \; a.
\end{equation}

The angle $\theta$ changes with time because both the viewing angle
and the direction of $\bmath{v}$ change. In particular, $\bmath{v}$
may acquire a component that is perpendicular to the plane of
Fig.~\ref{cartoon}, so that it makes some angle $\xi$ with this
plane. Writing $\tp$ for the angle between the LOS and the projected
velocity $\bmath{\vp}$ (see Fig.~\ref{cartoon}), we have
\begin{equation}
\cos\theta(\tem) = \cos\tp(\tem) \, \cos\xi(\tem)
\end{equation}
and hence
\begin{equation}
\sin\theta \; \theta\p = \cos\xi \, \sin\tp \; \tp\p + \cos\tp \,
\sin\xi \; \xi\p.
\end{equation}
However, since this equation should be evaluated at $\tem^1$ and
since $\xi(\tem^1) = 0$ and $\theta(\tem^1) = \tp(\tem^1)$ by
construction, we can simply ignore the perpendicular component and
write $\theta\p(\tem^1) = \tp\p(\tem^1)$ [unless $\theta(\tem^1) = 0$,
in which case the second term of equation \eref{dzddt} is zero
anyway].

The change in $\tp$ is given by the angles $\phi$ and $\psi$ between
the dashed and solid lines in Fig.~\ref{cartoon}(b). Clearly, $\phi$
is the parallax angle for a baseline of $\sin\theta \, v \, (\tem^2 -
\tem^1)$, i.e.\ the transverse distance between C and D, while $\psi$
denotes the change of $\bmath{v}$'s direction in the plane of the
drawing. Hence we obtain
\begin{eqnarray}
\theta\p & = & \frac{\psi - \phi}{\tem^2 - \tem^1} = \sin\eta \,
  \cos\zeta \; \frac{a}{v} \; - \frac{\sin\theta
\; v}{\dpa(\tem, \chi)}\nonumber\\
& = & \sin\eta \, \cos\zeta \; \frac{a}{v} \; - \frac{1+z}{\dpa(\tobs,
  \chi)} \, \sin \theta \; v.
\end{eqnarray}
$\dpa$ is the parallax distance between two objects separated by a
comoving coordinate distance $\chi$ \citep[e.g.][]{Peebles93,Peacock99}:
\begin{equation}
\dpa(t, \chi) = a(t) \frac{\Sigma(\chi)}{\Xi(\chi)}, 
\end{equation}
where we have used the notation
\begin{equation}
\Sigma(\chi) = \left\lbrace
\begin{array}{lccl}
\sin\chi \qquad & k & = & +1 \\
\chi & k & = & 0 \\
\sinh\chi & k & = & -1
\end{array} \right.
\end{equation}
and its cosine equivalent $\Xi(\chi) = \sqrt{1 - k \Sigma^2(\chi)}$.

\begin{figure}
\psfig{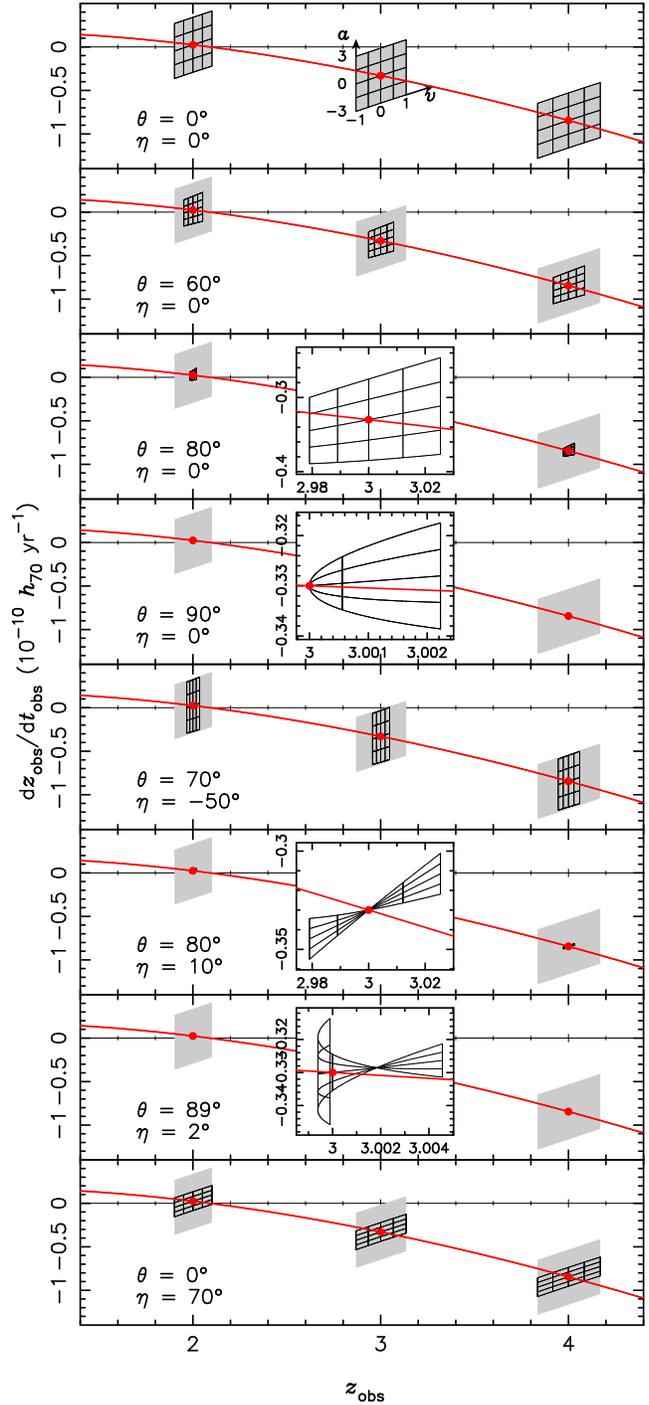}
\caption{Top panel: the solid line shows the cosmological redshift
  drift, $\dot z(z)$, expected for comoving sources (cf.\
  Fig.~\ref{dzdt}). The three grids show how an object at $z = 2$, $3$
  and $4$, respectively, moves in the observed $\zobs$-$\dot \zobs$
  plane as a function of $v$ and $a$, assuming $\theta = \eta = \zeta
  = 0$. For each grid, the vertical lines are lines of constant $v$,
  while the tilted horizontal lines are lines of constant $a$. The
  values of $v$ and $a$ are indicated along the middle grid in units
  of $10^4$\kms\ and $10^{-8}$~cm~s$^{-2}$. The other panels show the
  same for different values of $\theta$ and $\eta$ as indicated (and
  $\zeta = 0$). For comparison, the grey shaded areas in each panel
  are the same as in the top panel. The insets show a blow-up of the
  region around the $z = 3$ point.}
\label{dzodt}
\end{figure}

Collating all of the above we finally find for the observed redshift
drift:
\begin{eqnarray}
\label{dzobsdtobs}
\frac{\d \zobs}{\d \tobs} & = & \gamma \, \cos \zeta \, \cos \eta \; \ac
\left(\frac{\cos \theta}{1 + \cos \theta \vc} + \gamma^2 \vc \right)
\nonumber\\
& & \mbox{} - \gamma \, \cos \zeta \, \sin \eta \; \ac \; 
\frac{\sin \theta}{1 + \cos \theta \vc} \nonumber\\
& & \mbox{} + \gamma \; \frac{1+z}{\dpa} \; \frac{\sin^2\theta \,
  \frac{v^2}{c}}{1 + \cos \theta \vc} \nonumber\\
& & \mbox{} + (1 + \zobs) \, H(\tobs) - \gamma \, H(\tem) \nonumber\\
& = & (1 + \zobs) \, H_0 \nonumber\\
& & \mbox{} - \gamma \, \biggl \{ H(z) - \cos \zeta \; \ac 
\left[\frac{\cos(\theta + \eta)}{1 + \cos \theta \vc} 
+ \cos \eta \; \gamma^2 \vc\right]\biggr\} \nonumber\\
& & \mbox{} + \frac{1 + \zobs}{\dpa(z)} \; \frac{\vt^2}{c} \nonumber\\
& = &  (1 + \zobs) \, H_0 - \gamma \left[H(z) 
- \frac{\ar/c}{1 + \cos \theta \vc} 
- \gamma^2 \frac{v v\p}{c^2} \right] \nonumber\\
& & \mbox{} + \frac{1 + \zobs}{\dpa(z)} \; \frac{\vt^2}{c}
\end{eqnarray}
where we have used the notation
\begin{equation}
\vt = \frac{\sin \theta \; v}{1 + \cos \theta \vc}
\end{equation}
for the emitter's apparent transverse velocity and
\begin{equation}
\ar = \cos \zeta \, \cos(\theta + \eta) \; a
\end{equation}
for the magnitude of $\bmath{a}$'s component along the LOS.  Comparing
with equation \eref{zdot} we can see that the main difference is the
modification of $H(z)$ with a term proportional to the change in the
radial velocity. The second order term stems from taking the
derivative of $\gamma$ in equation \eref{dzddt}, while the last term
is due to the parallax effect. Clearly, equation \eref{dzobsdtobs}
reduces to equation \eref{zdot} for $v = a = 0$. Note also that
$\zeta$ is entirely degenerate with $a$ and so we will set $\zeta = 0$
from now on.

The formulae for many special cases such as $a = 0$ (constant motion),
$\theta = 0$ (radial motion) and $\theta = \pi/2$ (transverse motion)
are immediately apparent from equation \eref{dzobsdtobs}. Another
special case is that of a freely moving source. In the absence of
local gravitational potential wells we have $v \propto a^{-1}$ (and
$\zeta = \eta = 0$) so that $a = v\p = -v H(\tem)$. Neglecting terms
of order $(\vc)^4$ and higher, we obtain
\begin{equation}
\frac{\dot \zobs}{1 + \zobs} \approx 
H_0 - \frac{H(z)}{1 + z} \, \bigl(1 + \frac{\vt^2}{c^2}\bigr)
+ \frac{1}{\dpa(z)} \frac{\vt^2}{c},
\end{equation}
which shows that in this case purely radial peculiar motion preserves
the form of equation \eref{zdot}. Setting $\ol = 0$, we recover the
formula of \citet{Teuber86}:
\begin{equation}
\frac{\dot \zobs}{1 + \zobs} \approx 
H_0 \bigr[1  - \sqrt{1 + \om z} \; \bigl(1 + \frac{\vt^2}{c^2}\bigr)\bigr]
+ \frac{1}{\dpa(z)} \frac{\vt^2}{c}.
\end{equation}

Given an object's cosmological redshift $z$, and the parameters of its
peculiar motion ($v$, $\theta$) and ($a$, $\eta$) we can now predict
its observed position in a $\dot \zobs$ vs.\ $\zobs$ plot, i.e.\ the
observed equivalent of Fig.~\ref{dzdt}. This is shown in
Fig.~\ref{dzodt} for three objects at $z = 2$, $3$ and $4$,
respectively. The solid line in each panel shows the cosmological
$\dot z(z)$ that one expects in the absence of peculiar motions. In
each panel we fix $\theta$ and $\eta$ at some value and use small
grids to show how the three objects are perturbed away from their
`correct' positions (marked as dots) as a function of $v$ and $a$. The
grids cover the range $-1$ to $+1 \times 10^4$\kms\ in $v$ and $-3$ to
$+3 \times 10^{-8}$~cm~s$^{-2}$ in $a$. To relate these scales to one
another we point out that a constant acceleration of $1.5 \times
10^{-8}$~cm~s$^{-2}$ would result in a peculiar velocity of
$10^4$\kms\ by $z=3$.

Obviously, the largest possible effects are obtained when $\theta =
\eta = 0$ (top panel), and we reproduce this case as grey shaded areas
in all other panels for comparison. The extent of the grids in the $v$
and $a$ directions can be scaled independently and almost arbitrarily
by appropriately adjusting $\theta$ and $\eta$. However, when
$\theta+\eta \approx \pi/2$ purely relativistic and/or second order
effects take over. These are a factor of $\sim$$10$--$100$ smaller in
magnitude and can significantly distort the grids. Note that since the
grids are generally not symmetric, it is in principle possible
(contrary to our initial expectation) that peculiar motions could bias
a $\dot z(z)$ measurement, even when using a large sample of objects
whose peculiar motions are uncorrelated.

Given typical values for the peculiar velocities and accelerations of
a particular set of objects (e.g.\ Figs.~\ref{vp} and \ref{ap}), we
can essentially read off Fig.~\ref{dzodt} the scale of the noise on a
$\dot z(z)$ measurement introduced by the peculiar motions. This must
be compared to the photon noise on $\dot z(z)$ for an {\em individual}
object. In the case of the \lya\ forest, as discussed in this paper,
the photon noise error on \zdot\ for an individual absorption line is
about four orders of magnitude larger than the error induced by
peculiar motions of the absorbing gas. We conclude that peculiar
motions will only be relevant for a \zdot\ experiment if it is based
on very precise measurements on a small number of objects with large
peculiar velocities and accelerations.

\section{Summary of selected alternative cosmological models} 
\label{altcos}
There exist a number of cosmological models that may offer an
alternative to the standard \lcdm\ scenario. The purpose of this
Appendix is to provide a summary of the predictions for the redshift
drift of an incomplete subset of these alternative models. All of them
were introduced specifically to explain the observed acceleration of
the universal expansion, and they can be separated into two distinct
classes: (i) models that modify the stress-energy tensor of the
Universe to include a new dark energy component, but leave the general
relativistic field equation intact and (ii) models that modify the
theory of gravity.

Recall that the validity of the Robertson-Walker metric was the only
assumption we needed to derive equation \eref{zdot} for the redshift
drift (cf.\ Section \ref{dynamics}). None of the models considered in
this Appendix violate this assumption, and so all we need to do in
order to determine a given model's prediction for \zdot\ is to specify
the model's Friedman equation (see also \citealp*{Szydlowski06} for an
overview).

\subsection{Alternative dark energy models}
The general relativistic Friedman equation as given in equation
\eref{fried} assumes a specific form for each mass-energy component's
equation of state, namely $p_i = w_i c^2 \rho_i$, where $p_i$ and
$\rho_i$ are the component's pressure and density, respectively.
Without this assumption the Friedman equation reads:
\begin{equation}
\left[\frac{H(z)}{H_0}\right]^2 = \sum_i \Omega_i
\frac{\rho_i(z)}{\rho_{i0}} + \ok (1+z)^2,
\end{equation}
where $\rho_{i0}$ is the present-day value of $\rho_i$. For an
arbitrary equation of state, $p_i(\rho_i)$, the evolution
of $\rho_i$ is obtained by solving the general relativistic energy
conservation equation:
\begin{equation}
\frac{\d\rho_i}{\d a} = -\frac{3}{a} \, \left[\rho_i + 
\frac{p_i(\rho_i)}{c^2}\right].
\end{equation}

\subsubsection{Quintessence}
Here, we take the term quintessence to simply mean a smooth, time
varying dark energy component with equation of state $p_X = w_X c^2
\rho_X$, where $w_X$ may also vary with time. In principle, $w_X$ and
its evolution should be given by the scalar field equation of the
underlying physical theory of this component. In practice, however, it
is common to use a simple parameterisation of $w_X$ in order to be less
model dependent. Here we will use the expression of \citet{Linder03b}:
\begin{equation}
w_X(z) = w_0 + w_a \left(1 - \frac{a}{a_0}\right) = w_0 + w_a \frac{z}{1+z}.
\end{equation}
In this case the Friedman equation is given by
\begin{eqnarray}
\lefteqn{\left[\frac{H(z)}{H_0}\right]^2 = \Omega_X (1+z)^{3 (1+w_0+w_a)}
\, e^{-3 w_a z / (1+z)}} \nonumber\\
& & \mbox{} + \om (1+z)^3 + \ok (1+z)^2,
\end{eqnarray}
where we have included a matter component with $w_{\rm M} = 0$. Note
that for $(w_0, w_a) = (-1, 0)$ the above reduces to the standard
\lcdm\ case.

\subsubsection{Generalised Chaplygin gas}
In the \lcdm\ and quintessence models it is the different evolution of
the densities of the two mass-energy components that causes the
transition from a decelerating (matter-dominated) to an accelerating
($\Lambda$/quintessence-dominated) Universe. In contrast, the
generalised Chaplygin gas (GCG) model seeks to explain this transition
with only a {\em single} background fluid. This leads to a rather
exotic equation of state for GCG:
\begin{equation}
p_{\rm c} = -A \, c^2 \rho_{\rm c}^{-\alpha},
\end{equation}
where $A$ and $\alpha$ are constants with $A \ge 0$ and $\alpha \ge
-1$. Clearly, for $\alpha = -1$ GCG is identical to $\Lambda$ and
therefore the model is identical to the de Sitter Universe (containing
$\Lambda$ but no matter), while $\alpha = 1$ corresponds to the
standard Chaplygin gas model \citep{Kamenshchik01}. The above equation
of state results in the following Friedman equation:
\begin{eqnarray}
\lefteqn{\left[\frac{H(z)}{H_0}\right]^2 = \Omega_{\rm c} [B + (1 -
    B) (1+z)^{3(1+\alpha)}]^{1/(1+\alpha)}} \nonumber\\
& & \mbox{} + \ok (1+z)^2,
\end{eqnarray}
where $B = A/\rho_{{\rm c}0}^{1+\alpha}$. Note that we are neglecting
the baryons which had been subsumed in $\om$ previously. In this case
GCG is the only mass-energy component and we have $\Omega_{\rm c} = 1
- \ok$. Inserting the solution for $\rho_{\rm c}$ into the equation of
state we obtain:
\begin{equation}
p_{\rm c} = \frac{-B c^2}{B + (1-B)(1+z)^{3(1+\alpha)}} \; \rho_{\rm
  c} \equiv -w_{\rm c}(z) \, c^2 \rho_{\rm c}.
\end{equation}
From the above equations we can see clearly that for $0 < B < 1$ the
Hubble expansion and the GCG equation of state smoothly change from
being matter-like at early times to being $\Lambda$-like at late times
(possibly in the future), where the parameters $B$ and $\alpha$
jointly control when this transition occurs and how long it takes. For
the extreme values of $B = w_{\rm c}(0) = 0$ and $1$ GCG behaves like
pure matter and $\Lambda$, respectively, regardless of the value of
$\alpha$. Note that for $\alpha = 0$ the GCG model reduces to the
standard \lcdm\ case.

\subsection{Alternative gravity models}
\subsubsection{Cardassian models}
In Cardassian\footnote{With dark energy being `bad' \citep{White07}
and Cardassians being `ugly', all we need now is a `good' theory to
complete the famous trio.} models the Universe is assumed to be flat
and to contain only matter but no dark energy. The acceleration of the
expansion is instead achieved by altering the Friedman equation. In
the modified polytropic Cardassian models the Friedman equation reads
\citep{Wang03}:
\begin{equation}
\left[\frac{H(z)}{H_0}\right]^2 = \om (1+z)^3 [1 + (\om^{-q} - 1)
(1+z)^{3q(n-1)}]^{\frac{1}{q}},
\end{equation}
where the parameters $q > 0$ and $n < 2/3$ are constant. The case $q =
1$ corresponds to the original Cardassian model \citep{Freese02} which
is, however, as far as the Hubble expansion is concerned, exactly
equivalent to a flat quintessence model with $w_X = n-1$. Note that in
Cardassian models spatial flatness is {\em not} equivalent to $\om =
1$.

At early times the Cardassian Friedman equation is dominated by the
first term inside the brackets, resulting in normal, matter-dominated
expansion. However, at late times the non-standard term dominates and
gives rise to the acceleration. Such modifications to the Friedman
equation may be motivated by self-interacting dark matter or by
considering our observable Universe as a 3-dimensional brane
embedded in a higher dimensional space-time.

\subsubsection{DGP model}
Another brane-world scenario is the \citet*[DGP;][]{Dvali00} model. In
this (4+1)-dimensional model gravity can spread (or `leak') to the
flat, infinite extra spatial dimension at distances larger than some
cross-over scale $r_c$. On scales smaller than $r_c$ an observer on
the brane thus measures normal (3+1)-dimensional gravity, while on
larger scales gravity weakens, causing accelerated expansion beyond
$r_c$. If $r_c$ is chosen large enough then it will only enter the
Hubble horizon at late times so that the early evolution of the
Universe remains unaltered compared to standard, matter-dominated
general relativity, while at late times the leaking of gravity to the
extra dimension accelerates the expansion without requiring any form
of dark energy.

Defining an $r_c$-induced density parameter,
\begin{equation}
\Omega_{r_c} = \frac{c^2}{4 r_c^2 H_0^2},
\end{equation}
we can write the Friedman equation of the DGP model as
\citep{Deffayet02}:
\begin{eqnarray}
\lefteqn{\left[\frac{H(z)}{H_0}\right]^2 = \left[\sqrt{\Omega_{r_c}} +
  \sqrt{\Omega_{r_c} + \om (1+z)^3}\right]^2} \nonumber\\
& & \mbox{} + \ok (1+z)^2.
\end{eqnarray}
Note that in this model the curvature parameter is given by:
\begin{equation}
\ok = 1 - \left(\sqrt{\Omega_{r_c}} + \sqrt{\Omega_{r_c} + \om}\right)^2,
\end{equation}
so that flatness implies
\begin{equation}
\Omega_{r_c} = \left(\frac{1 - \om}{2}\right)^2.
\end{equation}

\label{lastpage}

\end{document}